\documentclass[
  journal=pasa,
  manuscript=article-type,
  year=2025,
  volume=37,
]{cup-journal}

\usepackage{amsmath}
\usepackage[nopatch]{microtype}
\usepackage{booktabs}
\usepackage{subcaption}

\title{On the origin of multifrequency temporal and spectral variability in Ton 599}

\author{Sakshi Maurya}
\affiliation{Department of Physics, Institute of Science, Banaras Hindu University, Varanasi-221005, India}

\author{Joysankar Majumdar}
\affiliation{Department of Physics, Institute of Science, Banaras Hindu University, Varanasi-221005, India}

\author{Varun}
\affiliation{School of Astronomy and Space Science, Nanjing University, Nanjing 210093, China}
\alsoaffiliation{Key Laboratory of Modern Astronomy and Astrophysics (Nanjing University), Ministry of Education, China}

\author{Neetu Sahu}
\affiliation{Center For Basic Sciences, Pt. Ravisankar Shukla University, Raipur, Chhattisgarh, India}

\author{Raj Prince}
\affiliation{Department of Physics, Institute of Science, Banaras Hindu University, Varanasi-221005, India}
\email[Dr. Raj Prince]{priraj@bhu.ac.in}

\begin{document}

\begin{abstract}
In this work, we studied the broadband temporal and spectral properties of the flat-spectrum radio quasar (FSRQ) Ton 599. 
We collected the long-term data from Jan 2019 to August 2024 when the source was in a long flaring episode. We used the Bayesian block methodology to identify the various flux states, including three flares. The broadband fractional variability is estimated during two flaring states. The F$_{\text{var}}$ variation with respect to frequency shows a nearly double hump structure similar to broadband SED. The Power spectral density (PSD) shows a pink-noise kind of stochastic variability in the light curve and we do not see any break in the power spectrum, suggesting a much longer characteristic time scale is involved in gamma-ray variability. The flux distribution is well-fitted with a double log-normal flux distribution, suggesting the variability of non-linear in nature. The gamma-ray, optical, and X-ray emissions were found to be highly correlated with a zero time lag, suggesting a co-spatial origin of their emissions. We used the one-zone leptonic model to reproduce the broad-band spectrum in the energy range from IR to very high-energy gamma-ray. The increase in the magnetic field and the Doppler factor were found to be the main cause for high flux states.  The XMM-Newton spectra taken during one of the flaring durations exhibit a signature of thermal black body emission from the accretion disk, suggesting a possible disk-jet coupling. This has also been indicated by the gamma-ray flux distribution, which shows the distribution as non-linear in nature, which is mostly seen in galactic X-ray binaries or AGN where the accretion disk dominates the emission. 
%This possibility of disk-jet coupling will be explored in the coming works. 
\end{abstract}

%\noindent Lorem ipsum dolor sit amet, consectetur adipiscing elit, sed do eiusmod tempor incididunt ut labore et dolore magna aliqua. 

\section{Introduction}
Active galactic nuclei (AGNs) are the nuclei of galaxies hosting supermassive black holes (SMBHs) at the center, which are active, meaning mass is accreting onto the black hole. AGN consists of three main parts, which are a central SMBH, a thermal accretion disk, and the relativistic jets in pairs produced perpendicular to the disk plane. There is also observational evidence of gas clouds hanging above the accretion disk, known as the broad-line region (BLR), and the dusty molecules having a torus-like structure to hide the central part of the AGN. AGNs are randomly oriented in the Universe where the jet points randomly. AGNs with one of the jets pointing toward the Earth are known as blazars \citep{1995PASP..107..803U}. The jets are highly relativistic in nature, which boosts all the emissions produced along the jet axis. 
%Blazars, which include BL Lacertae objects (BLL) and flat-spectrum radio quasars (FSRQs), are a subclass of radio-loud active galactic nuclei (AGN) characterized by emission primarily from a relativistic jet that is oriented towards Earth. 
The observational results suggest that they emit across the entire electromagnetic spectrum, ranging from low-energy radio waves to very high-energy $\gamma$-rays \citep{1997ARA&A..35..445U, HOVATTA2019101541}. Some blazars were also found to be in a good temporal and spatial correlation with the neutrino emissions or events detected by the IceCube observatory \citep{2018Sci...361.1378I, 10.1093/mnras/stad3804}. 
Blazars exhibit a very strong flux variability on the timescale from minutes to years across the entire electromagnetic spectrum \citep{1997ARA&A..35..445U, 2007ApJ...664L..71A, 2013MNRAS.436.1530R, HOVATTA2019101541, 2022ApJ...927..214G}. The broadband information is used to construct the broadband spectral energy distribution (SED) which shows two broad emission components \citep{1980ApJ...235..386M,1981ApJ...243..700K,1996ApJ...463..444S,1998MNRAS.299..433F,2010ApJ...710.1271A}, with the low-energy component peaking in the optical to X-ray range, which is explained by synchrotron emission from relativistic electrons traveling in the magnetic fields of the jet \citep{Rawes_2015,1982ApJ...253...38U}. The high energy component peaks at MeV to TeV energy range and it is believed to be the result of inverse Compton (IC) scattering of low-energy photons by the relativistic electrons within the jet (synchrotron-self Compton; SSC; \citealt{2009ApJ...704...38S}) or outside the jets (external Compton; EC; \citealt{1992A&A...256L..27D, 1994ApJ...421..153S}). The observation suggests that the peak of the low and high energy components in the SED also changes from source to source and sometimes also with the various flux states of the source. 
 Blazars also have a sub-category of flat spectrum radio quasars (FSRQs) and BL Lacerate objects (BL Lacs) depending upon the presence and absence of broad emission lines in their optical spectra. 
 Further, based on the location of the synchrotron peak blazars are classified into different sub-classes \citep{1995ApJ...444..567P} such as low synchrotron peak blazar (LSP for peak frequency below 10$^{14}$ Hz), high synchrotron peak blazar (HSP for peak frequency above 10$^{15}$ Hz), and intermediate synchrotron peak blazar (ISP for peak frequency between  10$^{14}$ Hz to 10$^{15}$ Hz) \citep{Abdo_2010c}. Separately, the BL Lacs objects were also classified into various categories depending on the location of the synchrotron peak. The classes are low BL Lac (LBL for peak frequency below 10$^{14}$ Hz), intermediate BL Lacs (IBL for peak frequency between 10$^{14}$ - 10$^{15}$ Hz), high BL Lacs (HBL for peak frequency above 10$^{15}$ Hz), and extreme high BL Lacs (EHBL where peak frequency is above 10$^{16}$ Hz).

It has also been shown that the jet emission is not entirely produced in the jet but also has some influence by the accretion disk in some of the AGNs and blazars. In \citet{Chatterjee_2018}, authors have shown the possible accretion disk origin of X-ray variability based on the X-ray PSD breaks derived from the combination of soft-X-ray telescope (SXT), LAPXC, and Swift-XRT. A moderate and strong correlation of disk and jet power is also obtained for a sample of blazars in \citet{2022PhRvD.106f3001R, 2024MNRAS.535.3595R}, suggesting disk-jet coupling in blazar is more common than we think. Based on the gamma-ray power spectrum \citet{10.1093/mnras/stad3399} have shown a possible disk-jet connection in three bright blazars (S4 0954+65, PKS 0903-57, and 4C +01.02).
During the current flaring episode in Ton 599, we see a possible hint of disk-jet coupling, which has been explored in detail. 

In section 2, we elaborate on the analysis procedure of various satellites. In section 3, we show our results and discuss them, followed by a summary in section 4.

\section{Multiwavelength Observations and Data Analysis}

 \subsection{Fermi-LAT}
Fermi-LAT is a $\gamma$-ray telescope launched by NASA in 2008 into a near Earth's orbit. It works using the pair-conversion method to detect high energy $\gamma$-rays between the energy range of 20 MeV to a Few hundred of GeV. When a gamma ray enters the telescope, it interacts with the material in the detector, and if the photon has sufficient energy, it may convert into an electron-positron pair through a process called pair production. A colorimeter at the base of the detector is placed to record the energies and the direction of these charged particles.
It is sensitive to photon energies between 20 MeV to higher than 500 GeV and has a field of view of roughly 2.4 Sr \citep{Atwood_2009}. The LAT's field of view covers approximately 20\% of the sky at any given moment, allowing it to scan the entire sky every three hours. Ton 599 has been continuously monitored by the Fermi-LAT since 2008. We analysed Fermi-LAT data from June 2020 to August 2024. The data is downloaded from the archive for a particular duration of time, and a circular region of 20 degrees is chosen around the source.
We followed the standard analysis procedure of {\tt Fermi Tools} and used {\tt Fermipy} to derive the light curve and the spectrum. We followed the documentation of the {\tt Fermipy}\footnote{\url{https://fermipy.readthedocs.io/en/latest/}}. To account for the galactic and extragalactic background, we have used the background files (gll\_iem\_v07.fits, iso\_P8R3\_SOURCE\_V3\_v1.txt) provided by the Fermi support team\footnote{\url{https://fermi.gsfc.nasa.gov/ssc/data/access/lat/BackgroundModels.html}}.

We produced the light curve for June 2020 to August 2024 and applied the Bayesian block method \citep{2022icrc.confE.868W} to identify the flares.  Three major flares were identified, and the remaining time source was in a quiescent state. We named them Flare 1,2, and 3. Flare 1 was observed during MJD 59370-MJD 59430, Flare 2 was observed during MJD 59440-MJD 59500, and Flare 3 was observed during MJD 59920-MJD 60030.
Maximum recorded flux during the Flare 1 was 2.31$\times$ 10$^{-6}$ ph cm$^{-2}$ s$^{-1}$, during the Flare 2 was 3.06$\times$ 10$^{-6}$ ph cm$^{-2}$ s$^{-1}$ and during the Flare 3 was 3.63$\times$10$^{-6}$ ph cm$^{-2}$ s$^{-1}$.
\begin{figure*}
    \centering
    \includegraphics[width=1.0\textwidth]{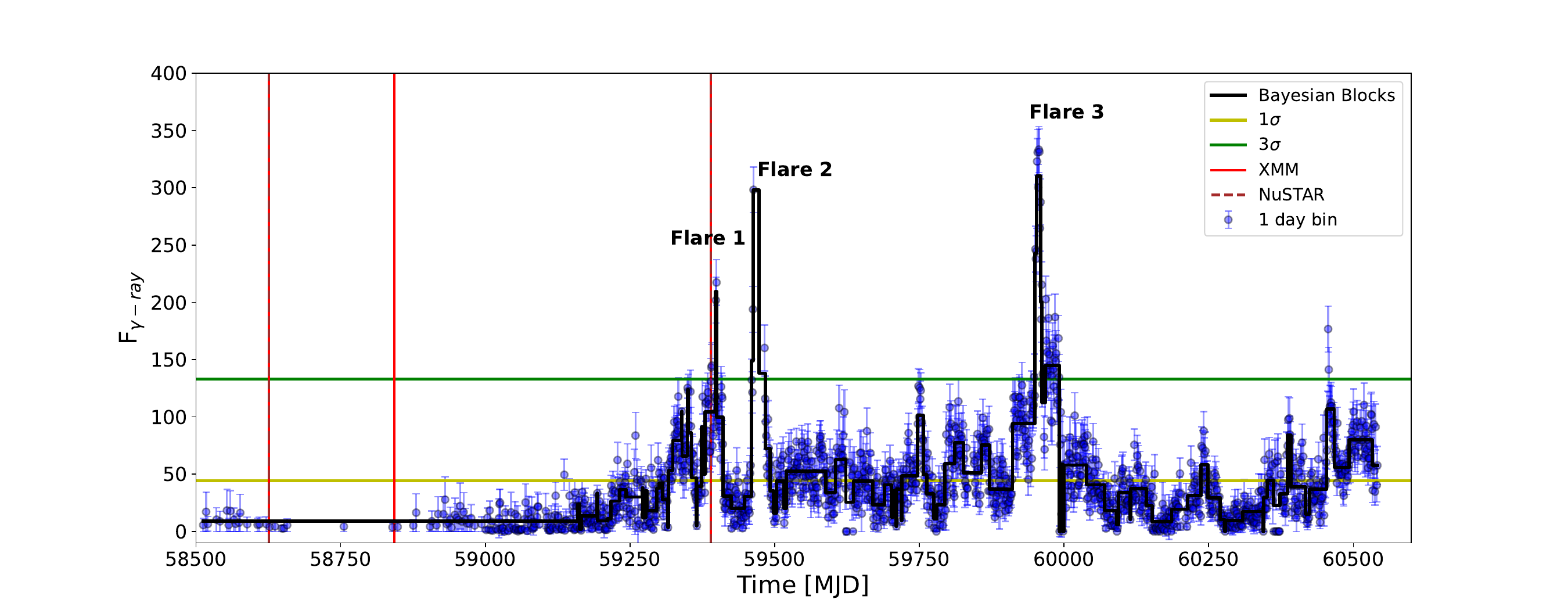}
    \caption{The $\gamma$-ray light curve using  1-day binning. Multiple flares have been identified by applying the Bayesian block method.}
    \label{fig_totallc}
\end{figure*}

\begin{figure*}
    \centering
    \includegraphics[width=1.0\textwidth]{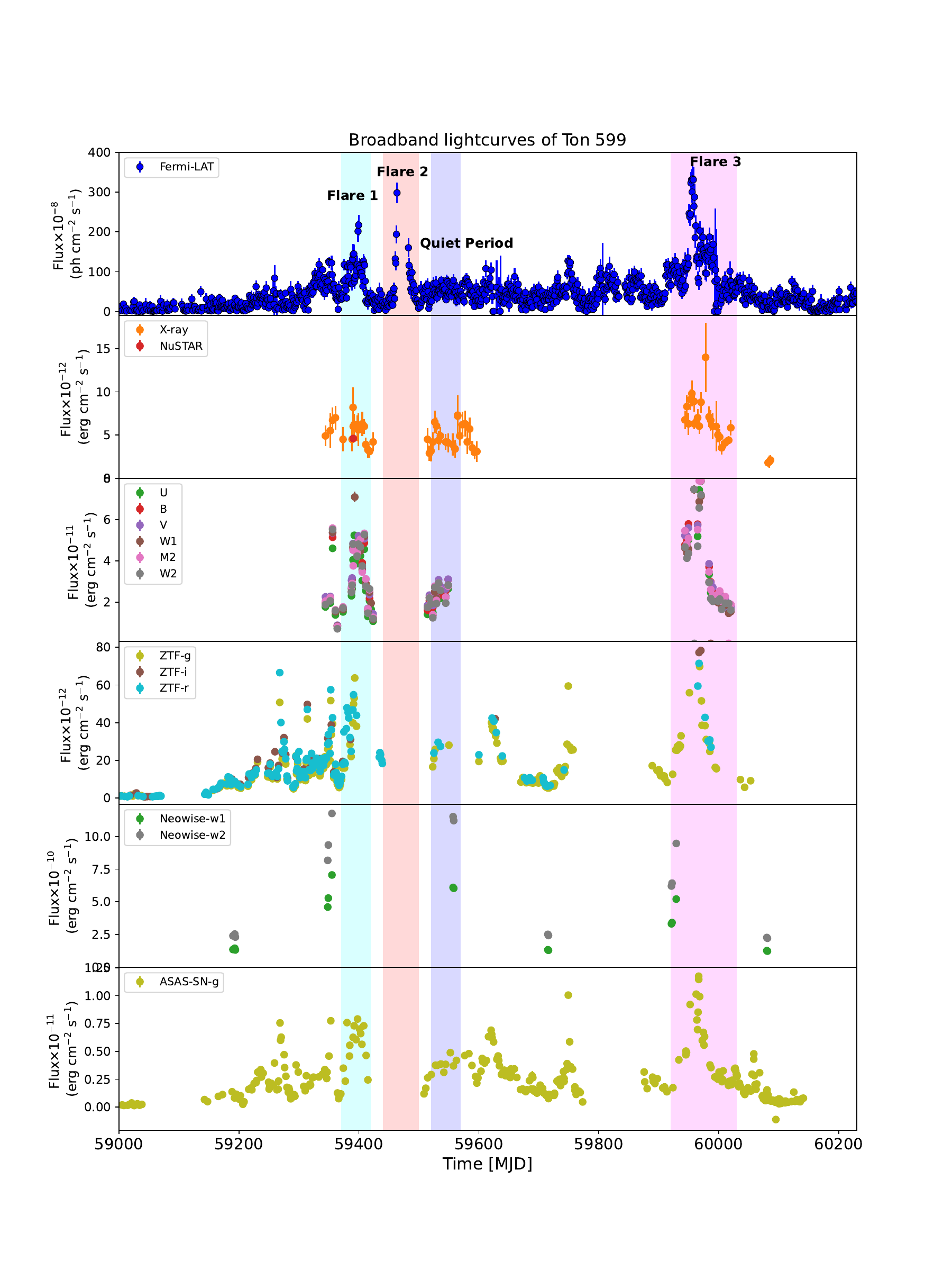}
    \caption{Broadband light curves of Ton 599 during the flaring episodes. The top panel shows the Fermi-LAT light curve, followed by the X-ray and UVOT light curves in panels 2 and 3. The archival ZTF, WISE, and ASAS-SN light curves are shown in panels 4,5, and 6.}
    \label{fig_lc}
\end{figure*}

\subsection{Swift-XRT/UVOT}
  Neil Gehrels Swift Observatory is a space-based observatory launched by NASA in 2004 to catch the brightest gamma-ray bursts (GRBs) in the sky. It consists of three main telescopes, namely: X-ray Telescope (XRT), Ultra-violet and Optical Telescope (UVOT), and Burst Alert Telescope (BAT). XRT and BAT are X-ray telescopes working in the energy range 0.3-10  (soft) keV and 3-150 (hard) keV, respectively. Most of the bright Fermi blazars detected in gamma rays have been monitored with the Swift\footnote{\url{https://www.swift.psu.edu/monitoring/}}. The advantage of Swift is that it can provide simultaneous observations of any object in soft and hard X-ray along with optical-UV. Particularly, the simultaneous observations across the wavebands are important in order to understand the physical mechanism producing occasional broadband spectacular flares in blazars. Blazar Ton 599 has been observed with Swift on multiple occasions during the gamma-ray flaring as well as non-flaring states. We have analysed all the swift observations available between May 2020 to September 2024. We present the outcome in this paper (see Table 1).  

 {\bf XRT:}
  We first downloaded the raw data from the HEASARC Archives and ran the {\tt XRTPIPELINE} to create the cleaned events files. Most of the observations are done with the photon counting (PC) mode, and hence, we took the cleaned event files of the PC mode to create the source and background regions. We chose the 20 arcsec radius around the source and away from the source for source and background regions and extracted the source and background spectra using the tool {\tt XSELECT}. We then collected the redistribution matrix file (RMF) from the command 'quzcif' using the {\tt XRTMKARF}; we then created the ancillary response file. All these files are loaded in {\tt GRPPHA} to merge and then are used in {\tt XSPEC} \citep{1996ASPC..101...17A} for modeling. We chose the simple power law to model the X-ray spectrum and also consider the galactic absorption ($N_H$=1.63$\times$10$^{20}$ cm$^{-2}$; \citealt{2016A&A...594A.116H}) and estimated the spectral index and the flux in the 0.3-10.0 keV energy regime. The estimated photon index and the flux are quoted in Table 1, along with its error bars. To create an average spectrum for the SED modeling, we add all the observations together during that particular period using {\tt ADDSPEC} and then model them in {\tt XSPEC} to derive the SED data points.

  {\bf UVOT:}
  During Swift's observation, the UVOT Telescope \citep{roming2005swift} observed the TON 599 in its three optical (U, B, and V) as well as in three ultraviolet(UVW1, UVM2, and UVW2) filters. We have downloaded the UVOT data from HEASARC Data archive\footnote{\url{https://heasarc.gsfc.nasa.gov/cgi-bin/W3Browse/w3browse.pl}} and performed the data analysis. We begin by accessing the {\tt uvot/image} within the specified {\tt obsid}, where images for all wavebands are located. Opening the filename-sk.img.gz file in DS9, we use {\tt SIMBAD} for object identification to generate the source and background region file. We then run {\tt uvotsource} for each filter separately to obtain AB magnitude. The source magnitudes were obtained from a circular region with a 5 arcsec radius centered on the source, while the background magnitudes were measured from a nearby, source-free circular region with a radius of 20 arcsec. For SED modeling, we summed up all the observation IDs during the particular period using {\tt uvotimsum} and extracted the source magnitude using {\tt uvotsource}. Magnitude is corrected for galactic extinction \citep{schlafly2011measuring} and converted into flux using zero-points and conversion factors \citep{larionov2016exceptional}.

\begin{table*}
   \caption{Log of the Observations during the Flaring State (MJD 59436-59509)}
   \label{KapSou2}
   \centering
%   \begin{array}{p{0.2\linewidth}p{0.16\linewidth}p{0.1\linewidth}p{0.1\linewidth}p{0.1\linewidth}p{0.1\linewidth}p{0.1\linewidth}}
    \begin{tabular}{ccccccc}
    \hline \hline
     \multicolumn{7}{c}{\textbf{Flare 1}} \\
      \hline \noalign{\smallskip}
      \textbf{Observatory} & \textbf{Obs-ID} & \textbf{Flux} & \textbf{Flux\_err} & \textbf{Photon} & \textbf{Photon} & \textbf{Exposure(ks)} \\
      & & [10$^{-12}$ erg cm$^{-2}$ s$^{-1}$] & [10$^{-12}$ erg cm$^{-2}$ s$^{-1}$] & \textbf{Index} & \textbf{Index-err} & \\
      \noalign{\smallskip}
      \hline
      \noalign{\smallskip}
      Swift-XRT/UVOT & 00036381058 & 4.50	& 1.40 & 1.54 & 0.29 & 0.72 \\
      Swift-XRT/UVOT & 00036381061 & 4.50	& 1.40 & 1.74 & 0.40 & 0.46 \\
      Swift-XRT/UVOT & 00036381062 & 8.20	& 2.30 & 0.98 & 0.38 & 0.78\\
      Swift-XRT/UVOT & 00088361002 & 6.00	& 0.70 & 1.48 & 0.16 & 4.55 \\
      Swift-XRT/UVOT & 00088361003 & 6.23 & 1.21 & 1.76 & 0.28 & 1.47 \\
      %Swift-XRT/UVOT & 00036381063 & 7.0	&1.0	&1.80	&0.17 & 1.53 \\
      Swift-XRT/UVOT & 00036381064 & 5.80 & 1.50 & 1.80 & 0.40 & 0.93 \\
      Swift-XRT/UVOT & 00036381065 & 5.60	& 1.90 & 1.90 & 0.50 & 0.63 \\
      Swift-XRT/UVOT & 00036381066 & 5.88 & 0.50 & 1.49 & 0.35 & 1.12 \\
      Swift-XRT/UVOT & 00036381067 & 6.30	&1.40&1.59&0.32 & 1.09 \\
      Swift-XRT/UVOT & 00036381068 & 6.00	&0.84&1.79&0.32 & 1.10 \\
      Swift-XRT/UVOT & 00036381069 & 3.90 &0.80&1.78&0.29 & 1.11 \\
      Swift-XRT/UVOT & 00036381070 & 3.30&0.90&1.22&0.35 &  1.12 \\
      Swift-XRT/UVOT & 00036381071 & 3.10	&0.70&1.86&0.45 & 1.05 \\
      %Swift-XRT/UVOT & 00036381072 & 4.77 &0.94	&1.80	&0.29 & 0.21 \\
      \noalign{\smallskip}
      \hline
      \multicolumn{7}{c}{\textbf{Flare 3}} \\
      \hline
      \noalign{\smallskip}
      Swift-XRT/UVOT & 00036381097 & 6.76	& 1.07 &1.80  &	0.16 & 1.40 \\
      Swift-XRT/UVOT & 00036381098 & 8.30	& 1.30 &1.49 & 0.19 & 1.48 \\
      Swift-XRT/UVOT & 00036381099 & 6.30	&1.30	&1.59	&0.22 & 0.98 \\
      Swift-XRT/UVOT & 00036381100 & 9.10	&1.2	&1.70	&0.14 & 1.57 \\
      Swift-XRT/UVOT & 00036381101 & 9.80	&1.50	&1.57	&0.14 & 1.59\\
      Swift-XRT/UVOT & 00036381102 & 8.90	&1.2	&1.57	&0.20 & 1.77 \\
      Swift-XRT/UVOT & 00036381103 & 6.30	&0.60	&1.60	&0.10 & 3.82 \\
      Swift-XRT/UVOT & 00036381104 & 7.0	&1.0	&1.80	&0.17 & 1.53 \\
      Swift-XRT/UVOT & 00036381105 & 6.03	&0.93	&1.85	&0.21 & 1.62 \\
      Swift-XRT/UVOT & 00036381106 & 8.80 & 1.20 & 1.42 & 0.14 & 1.67 \\
      Swift-XRT/UVOT & 00036381107 & 14.0	&4.0	&1.16	&0.20 & 0.39 \\
      Swift-XRT/UVOT & 00036381109 & 7.10	&1.20	&1.64	&0.20 & 1.44 \\
      Swift-XRT/UVOT & 00036381110 & 6.70	&1.20	&1.57	&0.28 & 1.39 \\
      Swift-XRT/UVOT & 00036381111 & 6.00	&1.50	&1.46	&0.19 & 1.27 \\
      Swift-XRT/UVOT & 00036381113 & 6.0	&2.9	&1.4	&0.6 & 0.28 \\
      Swift-XRT/UVOT & 00036381114 & 5.02	&0.88	&1.58	&0.19 & 1.34 \\
      Swift-XRT/UVOT & 00036381115 & 4.77 &0.94	&1.80	&0.29 & 1.26 \\
      Swift-XRT/UVOT & 00036381116 & 3.52	&0.81	&2.0	&0.40 & 0.97 \\
      Swift-XRT/UVOT & 00036381117 & 4.10	&0.60	&2.07	&0.19 & 1.28 \\
      Swift-XRT/UVOT & 00036381121 & 4.40	&0.5	&1.44	&0.21 & 1.30 \\
      Swift-XRT/UVOT & 00036381122 & 5.84	&0.87	&1.78	&0.12 & 1.36 \\
      \noalign{\smallskip}
      \hline
   \end{tabular}
\end{table*}

 \subsection{NuSTAR}
The Nuclear Spectroscopic Telescope Array (NuSTAR) mission, launched on 2012 June 13, is the first focusing high-energy X-ray telescope in orbit \citep{harrison2013nuclear}. It consists of two co-aligned X-ray detectors paired with their corresponding focal plane modules, called FPMA and FPMB, and it operates over a wide energy range from 3 to 79 keV.  It recorded two observations for Ton 599; we have analysed one observation (Obsid 60463037004) for 2021-06-25 (MJD 59390.53). We have downloaded the NuSTAR data from the HEASARC Data archive and ran the {\tt nupipeline}, so two cleaned event files were produced. We extracted the source and background spectra using a circular region and used the tool {\tt nuproducts}. We grouped the spectra using {\tt grppha} with 30 counts per bin and then used {\tt XSPEC} for modeling.
\subsection{XMM-Newton}
In gamma-ray, the source was reported to be flaring during June 2021 through a telegram (ATel\#14722; \citealt{2021ATel14722....1P}). Following this event, we proposed a target-of-opportunity observation in XMM-Newton to study the short-term variability in X-rays. Finally, the observation was done on 2021-06-25 at 18:52:30.
We also looked at the archive and collected the observations done during a low flux state. The timeline of the observations is marked in Figure 1.

We followed the standard analysis procedure to analyse the XMM-Newton data using the Science Analysis Software (SAS) version 18.0.0. We extracted both the light curves and the spectra by selecting a source region of 20 arcsecs around the source and a background region of a circular radius of 50 arcsecs away from the source. The details of the observations are shown in Table \ref{tab:xmm_nustar}.

\begin{table}
\centering
\renewcommand{\arraystretch}{1.2}
\caption{The observational log of XMM-Newton and NuSTAR, which are used in this work.}
\begin{tabular} {cccc}
\hline \hline
\noalign{\smallskip}
Observatory & Obs-ID & Date of  & Exposure \\
 & & Observation & time (ks) \\
\hline \noalign{\smallskip} 
XMM-Newton& 0850390101 & 2019-05-23&  18 \\
XMM-Newton & 0850390102 & 2019-12-26 & 13  \\
XMM-Newton & 0850390103 & 2021-06-25& $\sim$ 15  \\
\hline \noalign{\smallskip}
NuSTAR  & 60463037002 & 2019-05-23 	& 17 \\
NuSTAR  & 60463037004 & 2021-06-25  & $\sim$18 \\
\noalign{\smallskip} \hline
\end{tabular}
\label{tab:xmm_nustar}
\end{table}

\subsection{ZTF}
Zwicky Transient Facility (ZTF) is a new optical time-domain survey that uses the Palomar 48-inch Schmidt telescope. ZTF uses a 47 $deg^2$ field with a 600-megapixel camera to scan the entire northern visible sky at rates of $\sim 3760$ $deg^2$ per hour \citep{bellm2018zwicky}. ZTF has three filters: ZTF-g, ZTF-i, and ZTF-r. We have downloaded ZTF observations from the NASA/IPAC Infrared Science Archive\footnote{\url{https://irsa.ipac.caltech.edu/frontpage/}}. We get the magnitude(AB) and time (MJD), convert the magnitude into flux using zero-points, and take the center wavelength $\lambda_{\text{cen}}$ from SVO filter service\footnote{\url{http://svo2.cab.inta-csic.es/theory/fps/}}.

 \subsection{NEOWISE}
The Wide Field Infrared Explorer(WISE) has surveyed the entire sky in four infrared wavelengths with high sensitivity and spatial resolution devices. NEOWISE(Near-Earth Object Wide-Field Infrared Survey Explorer) is the enhancement of WISE. NEOWISE mission continues to detect, track, and characterise minor planets \citep{mainzer2011preliminary}. The four infrared filters are W1, W2, W3, and W4; we have taken only W1 and W2 in our study. We downloaded the data from the Infrared Science Archive and changed AB magnitude to flux using center wavelength $\lambda_{\text{cen}}$ from the SVO\footnote{\url{http://svo2.cab.inta-csic.es/theory/fps/}} filter service.
 \subsection{ASAS-SN}
 All-Sky Automated Survey for Supernovae (ASAS-SN), currently consisting of 24 telescopes around the globe, automatically surveys the entire visible sky every night down to about 18th magnitude in both V and g filters. We have accessed the data using ASAS-SN Sky Patrol V2.0\footnote{http://asas-sn.ifa.hawaii.edu/skypatrol} \citep{2023arXiv230403791H,2014ApJ...788...48S}. We took only the g-filter data with good(G) quality, used it in the light curve, and took the average flux for the particular period to use it in SED modeling.
 
 \subsection{Archival}
 We have downloaded the archival data from SED Builder\footnote{\url{https://tools.ssdc.asi.it/SED/}} and use it as background data in our broadband SED modeling. This helps us to guide the SED modeling and put better constraints on the total SED.

\section{Results and Discussions}
We have analysed Fermi-LAT, Swift-XRT/UVOT data from 2019 Jan to August 2024 Jan (MJD 58500—MJD 60542). We have also collected archival data to study the broadband behavior of the source.

\subsection{Gamma-ray ($\gamma-ray$) Light curve}
As shown in the light curve (Figure~\ref{fig_totallc}), three major flares are observed in Ton 599 from 2019 to 2024. We have studied the temporal evolution of all the flares separately. To illustrate the temporal evolution, we fitted the peaks with a sum of exponential functions, providing the rise and decay times for each peak visible in the light curve plots.  The functional form of the sum of exponentials is given by \cite{Abdo_2010}
\begin{equation}
F(t) = 2F_0 \left[ \exp\left( \frac{t_0 - t}{T_r} \right) + \exp\left( \frac{t - t_0}{T_d} \right) \right]^{-1}
\end{equation}

where $F_0$ is the flux at time $t_0$ representing the approximate flare amplitude, and $T_r$ and $T_d$ are the rise and decay times of the flare.

\begin{figure*}
    \centering
    \includegraphics[scale=0.37]{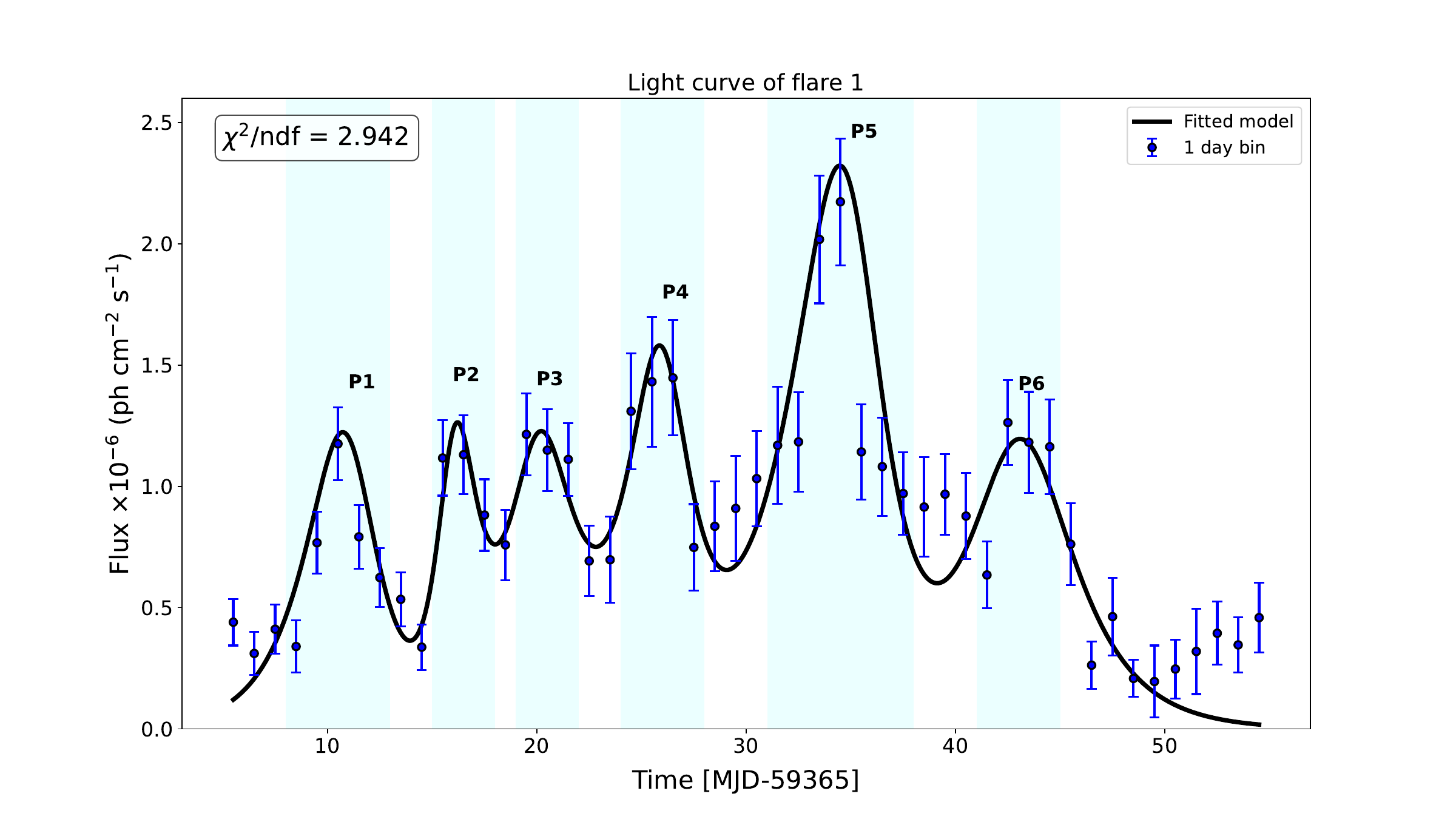}
    \includegraphics[scale=0.32]{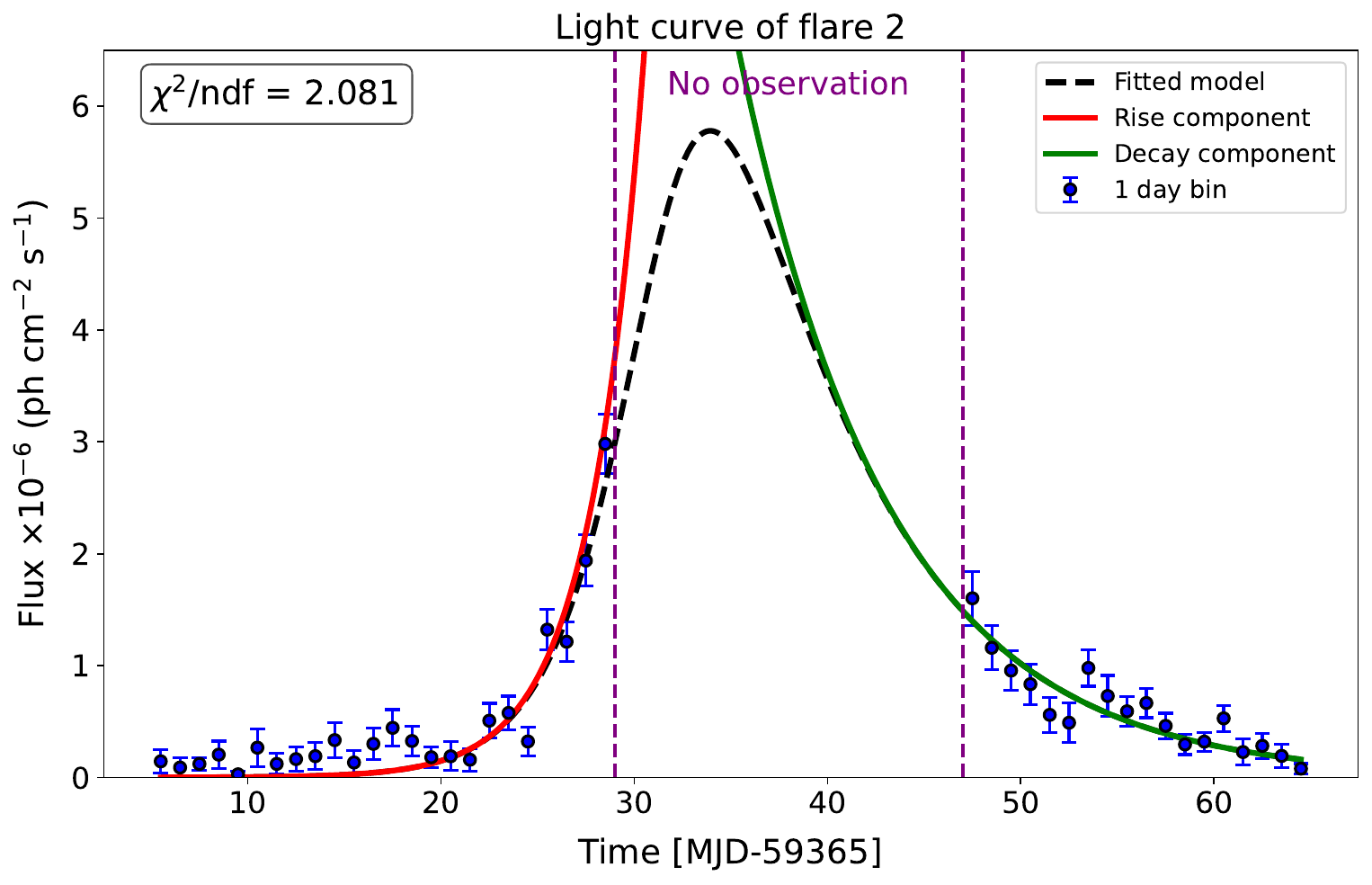}
    \includegraphics[scale=0.32]{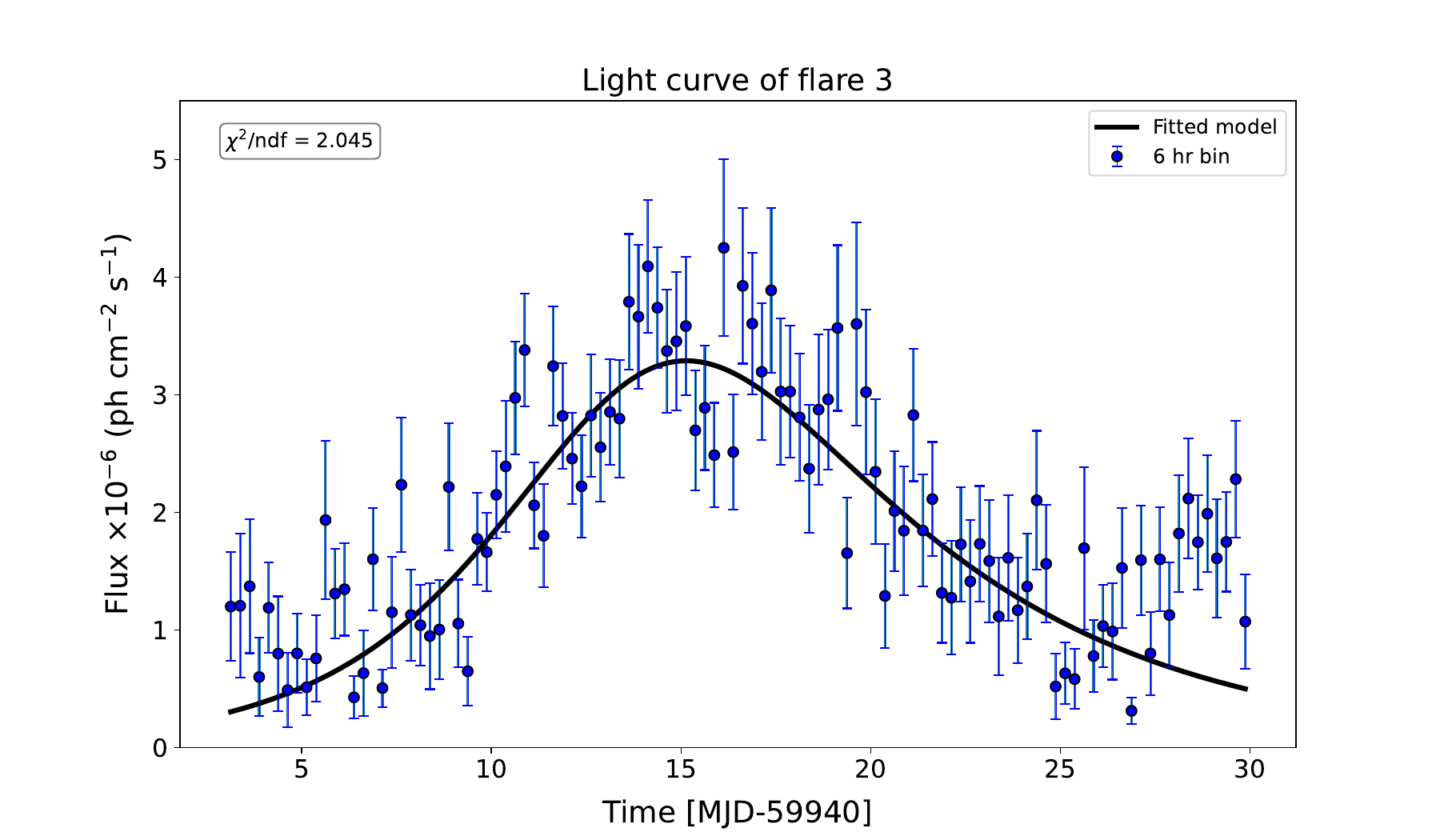}
        \caption{The local peak in the $\gamma$-ray light curve for each flare is fitted with the Sum of the exponential function. In flare 2, there was not sufficient observation, so we tried to fit it with the rising and decay function. The reduced $\chi^2/\text{ndf}$ values are calculated to estimate the goodness-of-fit and are mentioned in each plot. }
    \label{fig_rise-decay}
\end{figure*}

Figure~\ref{fig_rise-decay} shows the light curve of flare 1 and flare 2 with 1-day binning corresponding to the flaring activity during MJD 59370-59420 and 59440-59500. The flare 1 shows six peaks $P_1$, $P_2$, $P_3$, $P_4$, $P_5$ and $P_6$ and maximum flux during the flare 1 was 2.18$\times$ 10$^{-6}$ ph cm$^{-2}$ s$^{-1}$, which is the flux of peak $P_5$. In flare 2, no Fermi-LAT data is available in the time range MJD 59394-59412, so the sum of exponentials does not fit it. So, we have tried to fit it using the rise and decay function of the sum of exponentials. The maximum flux during flare 2 was 3.06$\times$ 10$^{-6}$ ph cm$^{-2}$ s$^{-1}$ but it is not fitted finely so that showing the highest flux 5.13$\pm$0.16$\times$ 10$^{-6}$ ph cm$^{-2}$ s$^{-1}$. The light curve of flare 3 with 6-hour binning corresponds to its flaring activity during MJD 59920-60030.

The result after fitting is shown in Table~\ref{KapSou0}, where rise time $T_r$ and decay time $T_d$ for each flare are mentioned. Flare 1 shows six bright peaks, and their rise and decay time varies from 1.06 to 2.35 hours. Peak $P_2$, $P_3$, $P_4$, and $P_6$ is symmetric in nature with almost equal rise and decay time (within errorbars) suggesting it may occur when a perturbation in the jet flow or a blob of denser plasma passes through a standing shock present in the jet \citep{1979ApJ...232...34B}. At peak, $P_1$ and $P_5$ rise is comparatively slower than the decay, suggesting a slow injection of electrons into the emission region. In flare 2, the peak rises very fast with a rise time of around 2.78$\pm$0.12 hours, and the flux decays very slowly, with a decay time of 7.83$\pm$1.13 hours. The fast rise and slow decay in Flare 2 suggest a longer cooling time for the electrons through the various processes. A similar trend is also seen in flare 3, where the decay is slower compared to the rise, suggesting slower cooling of the electrons. It has been argued that any physical process faster than the light travel time will not be detectable from the light curve, and hence, the rise and decay time will always be higher than the light crossing time, and all three scenarios of rise and decay time are possible.

\subsection{Multi-wavelength light curves}
In Figure \ref{fig_lc}, we show the broadband light curves collected from various telescopes. In panel 1, we present a 1-day binned gamma-ray light curve where various flares and quiet periods have been marked in different patches of colours. During the flaring events, the object was also monitored in X-rays with Swift-XRT. In the second panel, we show the XRT light curve for energy 0.3-10 keV. We observed high flux variability during flare 3, but during flare 1 and the quiet period, the flux did not vary much. During flare 2, we do not have any X-ray observations. The flaring behavior in optical-UV is much clearer compared to X-ray, and the UVOT light curve is shown in panel 3. A close temporal correlation between the Fermi light curve and the UVOT light curve is seen, suggesting that gamma-ray and optical emissions are highly correlated and produced at the same time. A detailed study of the correlation is presented later in this paper. The archival optical light curves from ZTF (g, i, r-bands) and ASAS-SN (g-band) are shown in panels 4 and 6. They clearly follow the UVOT light curve and are in temporal correlation with the gamma-ray emission. In panel 5, we show the archival near-infrared emission from the WISE observatory. Unfortunately, we do not have a clear temporal correlation of gamma-ray with WISE band emission, but we see some correlation with optical band emission. The overall suggestion is that the broadband light curves are correlated temporally and might have been produced at the same location and at the same time. This information is essential for broadband SED modeling in order to derive the main physical mechanism responsible for their emissions.

\subsection{Fractional Variability}
Blazars show strong flux variability at all frequencies. Fractional variability allows for the comparison of variability amplitudes across the entire electromagnetic spectrum and can be calculated using the relation given in \cite{vaughan2003characterizing}.
\begin{equation}
F_{\text{var}} = \sqrt{\frac{S^2 - \sigma^2}{r^2}},
\end{equation}

\begin{equation}
\text{err}(F_{\text{var}}) = \sqrt{\left(\sqrt{\frac{1}{2N}}\cdot \frac{\sigma^2}{ r^2 F_{\text{var}}} \right)^2 + \left( \sqrt{\frac{\sigma^2}{N}}\cdot\frac{1}{r} \right)^2},
\end{equation}

where $\sigma^2_{\text{XS}} = S^2 - \sigma^2$, is called excess variance, $S^2$ is the sample variance, $\sigma^2$ is the mean square uncertainties of each observation, and r is the sample mean. 

\begin{table*}
   \caption{Results of Temporal Fitting with Sum of Exponentials}
   \label{KapSou0}
   \centering
  %\begin{array}{p{0.2\linewidth}p{0.2\linewidth}p{0.2\linewidth}p{0.2\linewidth}p{0.2\linewidth}}
    \begin{tabular}{ p{3cm}p{3cm} p{3cm}p{3cm}p{3cm} }
      \hline \hline
       \multicolumn{5}{c}{\textbf{Flare 1}} \\
       \hline \noalign{\smallskip}
      \textbf{Peak} & \textbf{\boldmath{$t_0$}} & \textbf{\boldmath{$F_0$}} & \textbf{\boldmath{$T_r$}} & \textbf{\boldmath{$T_d$}}  \\
      &  [MJD] & [10$^{-6}$erg cm$^{-2}$ s$^{-1}$] & (hr) & (hr)  \\
      \noalign{\smallskip}
      \hline 
      \noalign{\smallskip}
      P1 & 59376 & 1.20 & 1.83$\pm$0.05 & 1.26$\pm$0.10 \\
      P2 & 59381 & 1.10 & 1.06$\pm$0.08 & 1.06$\pm$0.58 \\
      P3 & 59385 & 1.10 & 1.20$\pm$0.81 & 1.62$\pm$0.38 \\
      P4 & 59391 & 1.38 & 1.49$\pm$0.63 & 1.16$\pm$0.19 \\
      P5 & 59400 & 2.18 & 2.60$\pm$0.17 & 1.41$\pm$0.12 \\
      P6 & 59408 & 1.18 & 2.11$\pm$0.46 & 2.35$\pm$0.10 \\
     \noalign{\smallskip}
      \hline \hline
      \multicolumn{5}{c}{\textbf{Flare 2}} \\
      \hline 
      \noalign{\smallskip}
      P&59466&5.13$\pm$0.16&2.78$\pm$0.12&7.83$\pm$1.13 \\
      \noalign{\smallskip}
      \hline \hline
      \multicolumn{5}{c}{\textbf{Flare 3}} \\
      \hline \noalign{\smallskip}
        P & 59953 & 3.16$\pm$0.39	& 3.53$\pm$0.25  &6.33$\pm$0.66   \\
     \hline \noalign{\smallskip}
   \end{tabular}
\end{table*}

\begin{figure*}
    \centering
    \includegraphics[scale=0.37]{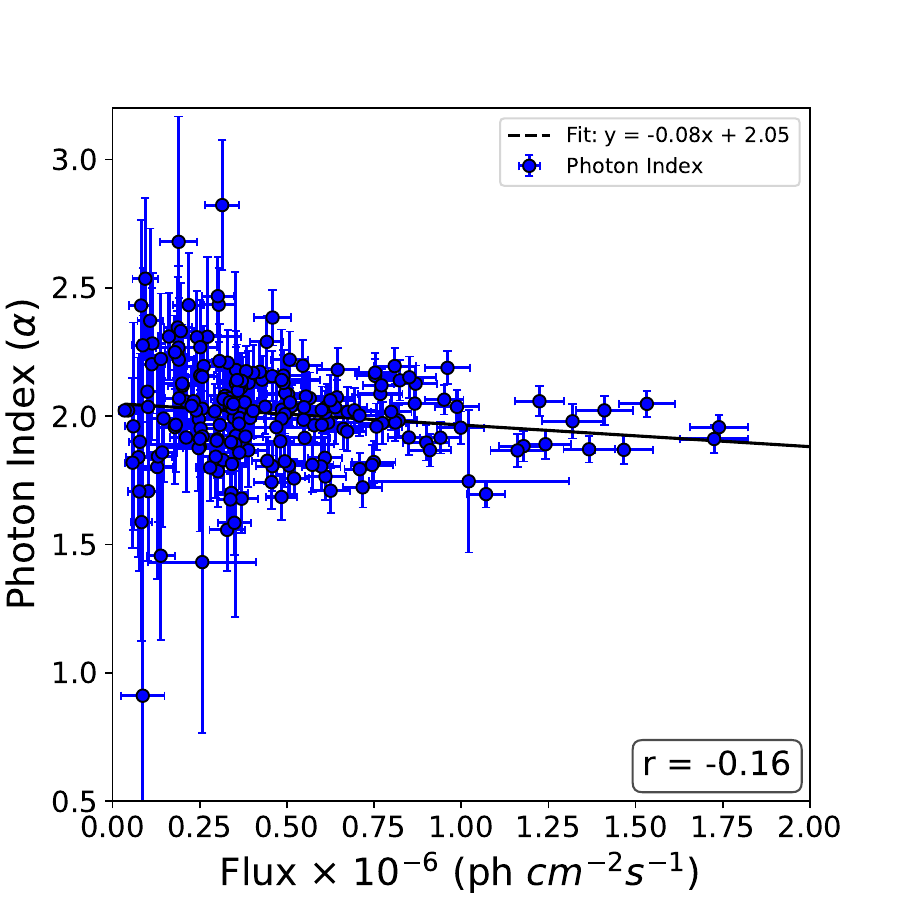}
    \includegraphics[scale=0.37]{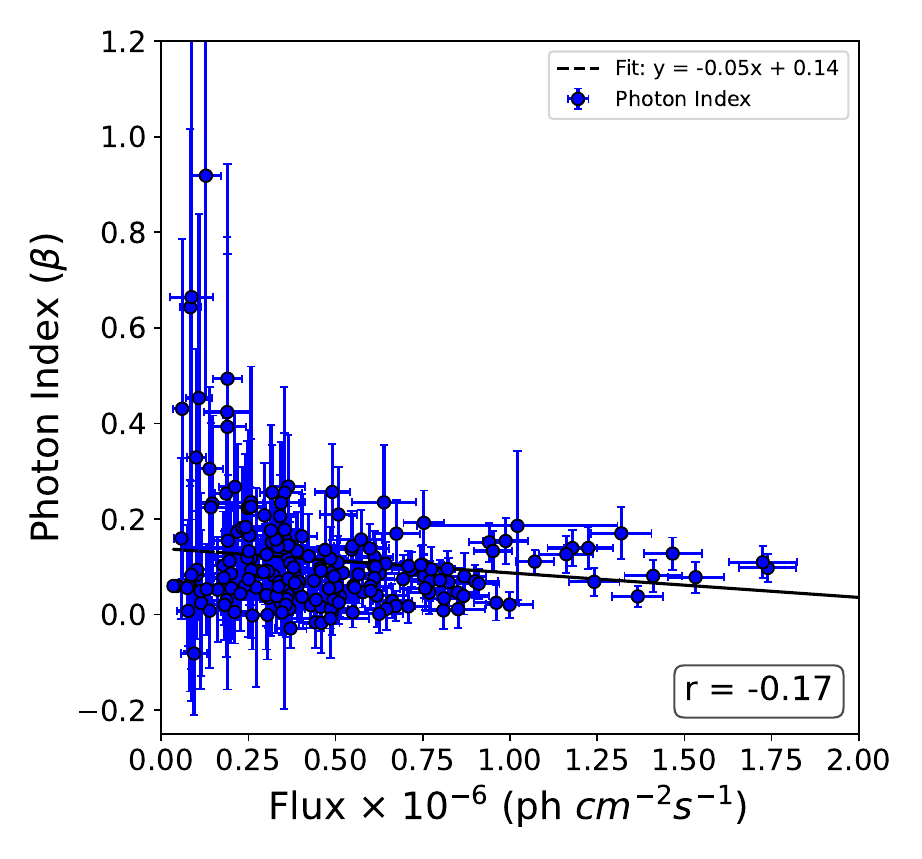}
    \includegraphics[scale=0.37]{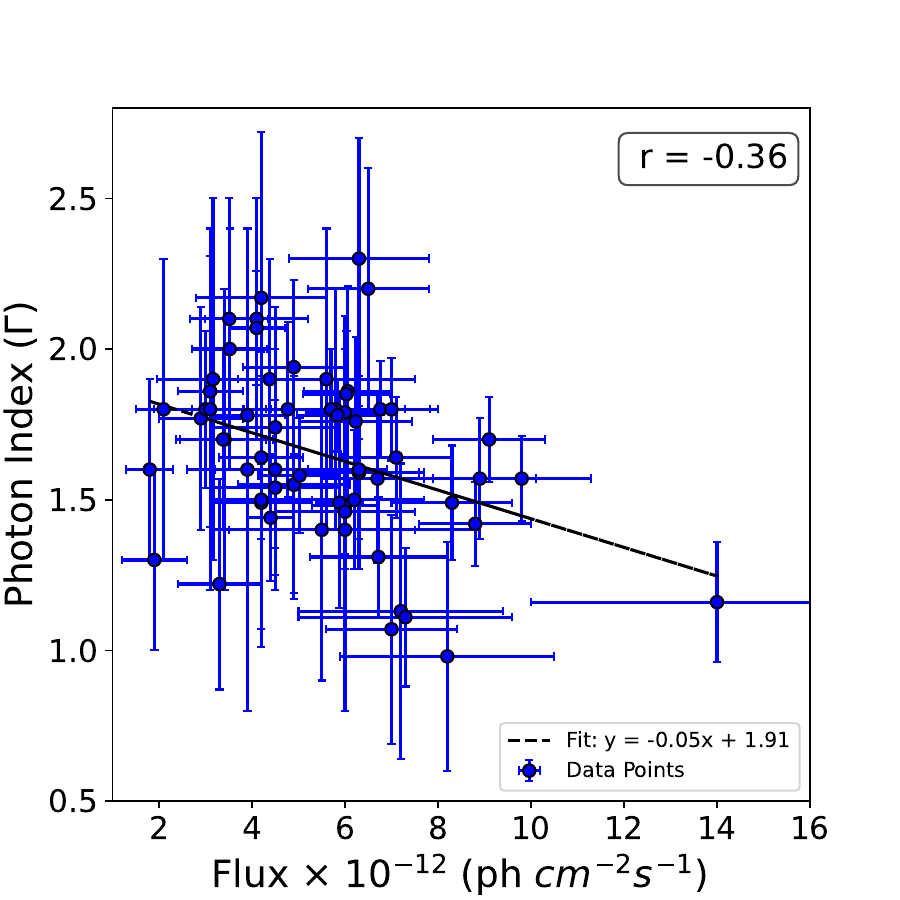}
                \caption{$\gamma$-ray Photon Index ($\alpha$) and curvature parameter ($\beta$) vs gamma-ray flux and X-ray photon index vs X-ray flux. }
    \label{fig_flux-index}
\end{figure*}

\begin{figure*}
    \centering
    \includegraphics[scale=0.295]{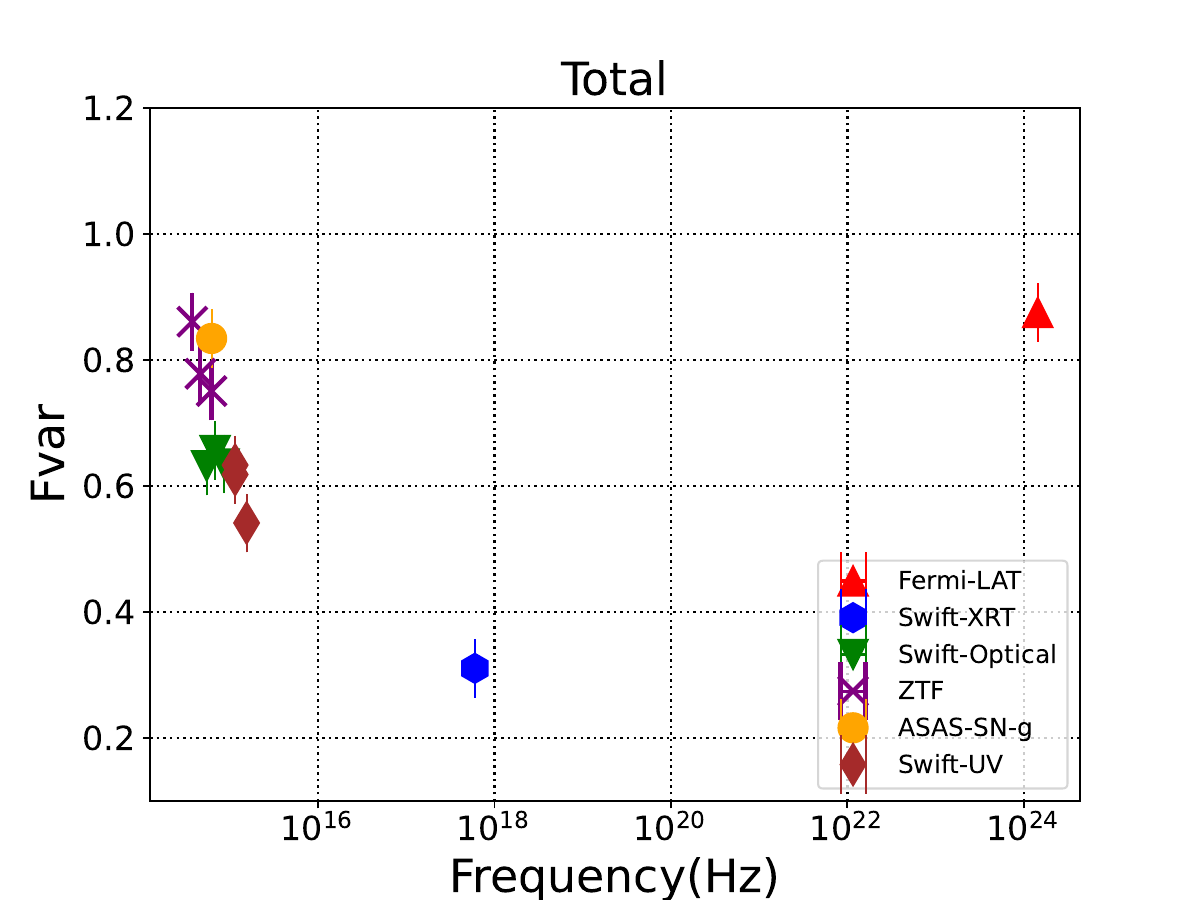}
    \includegraphics[scale=0.295]{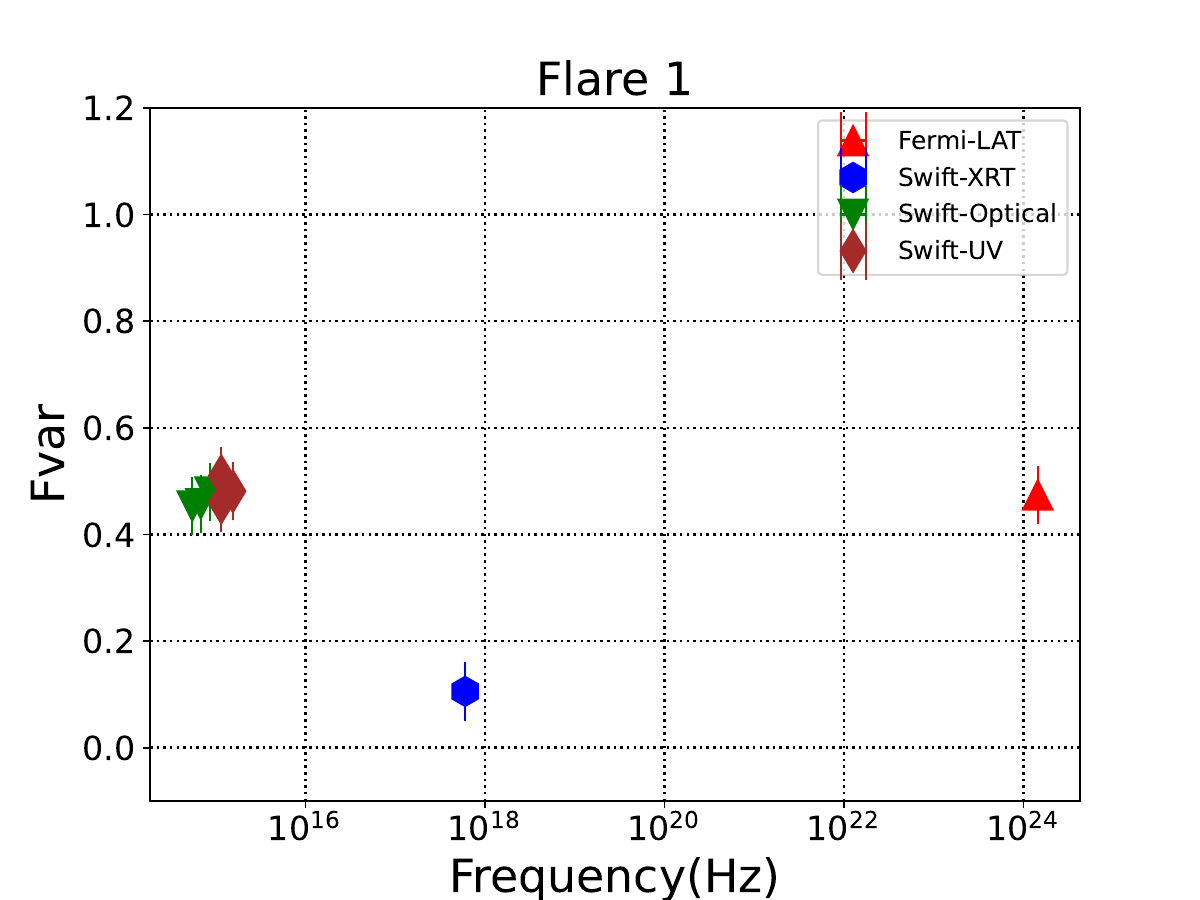}
    \includegraphics[scale=0.295]{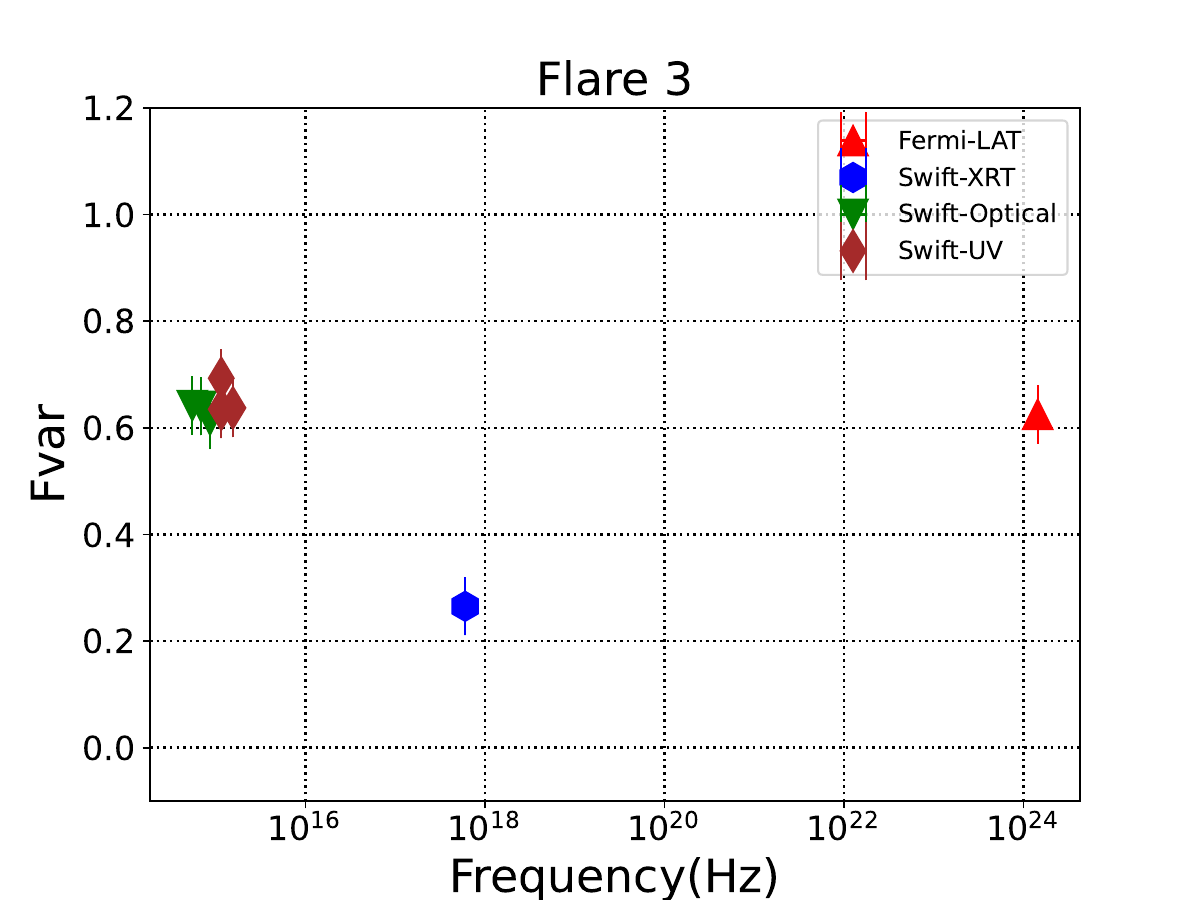}
    \caption{Fractional Variability estimated for various waveband for flaring state, flare 1, 2, and total dataset.}
    \label{fig:fig_fvar}
\end{figure*}

The plot of $F_{\text{var}}$ versus frequency is shown in Figure~\ref{fig:fig_fvar} to illustrate its overall shape. We have created plots for three segments total dataset, Flare 1 and Flare 3. However, due to the limited number of data points, Flare 2 has not been plotted. All segments show a similar shape; $\gamma$ -ray shows the highest variability, followed by optical and UV bands and then X-ray. A similar result is also obtained for other blazars by \cite{galaxies7020062}.
In literature, studies have also shown the increasing or decreasing F$_{var}$ with respect to frequency, suggesting either the large number of particles injected in the jet \citep{Prince_2020} or the presence of steady thermal emission \citep{Bonning_2009}.

\begin{table}
\centering
\renewcommand{\arraystretch}{1.5}
\caption{Fractional Variability}
\begin{tabular}{ p{1.8cm} p{1.8cm} p{1.8cm} p{1.8cm} }
\hline \hline
\noalign{\smallskip}
Wavebands & \multicolumn{3}{c}{F$_{\text{var}}$} \\
\hline
 & Total & Flare 1 & Flare 3 \\
\noalign{\smallskip}
\hline \noalign{\smallskip}
$\gamma$-ray & 0.87$\pm$0.01 & 0.47$\pm$0.03 & 0.62$\pm$0.03 \\
Swift-XRT & 0.31$\pm$0.03 & 0.10$\pm$0.01 & 0.26$\pm$0.05 \\
ZTF-g & 0.750$\pm$0.001 & - & - \\
ZTF-r & 0.860$\pm$0.002 & - & - \\
ZTF-i & 0.778$\pm$0.001 & - & - \\
ASAS-SN-g & 0.834$\pm$0.002 & - & - \\
U & 0.635$\pm$0.004 & 0.479$\pm$0.008 & 0.614$\pm$0.007 \\
B & 0.656$\pm$0.004 & 0.457$\pm$0.008 & 0.640$\pm$0.007 \\
V & 0.632$\pm$0.006 & 0.453$\pm$0.010 & 0.641$\pm$0.008 \\
W1 & 0.633$\pm$0.005 & 0.510$\pm$0.009 & 0.693$\pm$0.006 \\
M2 & 0.618$\pm$0.005 & 0.459$\pm$0.008 & 0.634$\pm$0.007 \\
W2 & 0.541$\pm$0.005 & 0.481$\pm$0.009 & 0.637$\pm$0.005 \\
\noalign{\smallskip}
\hline
\end{tabular}
\end{table}

\subsection{X-ray spectral fitting}
\begin{figure*}
    \centering
    \includegraphics[angle=-90, width=0.47\textwidth]{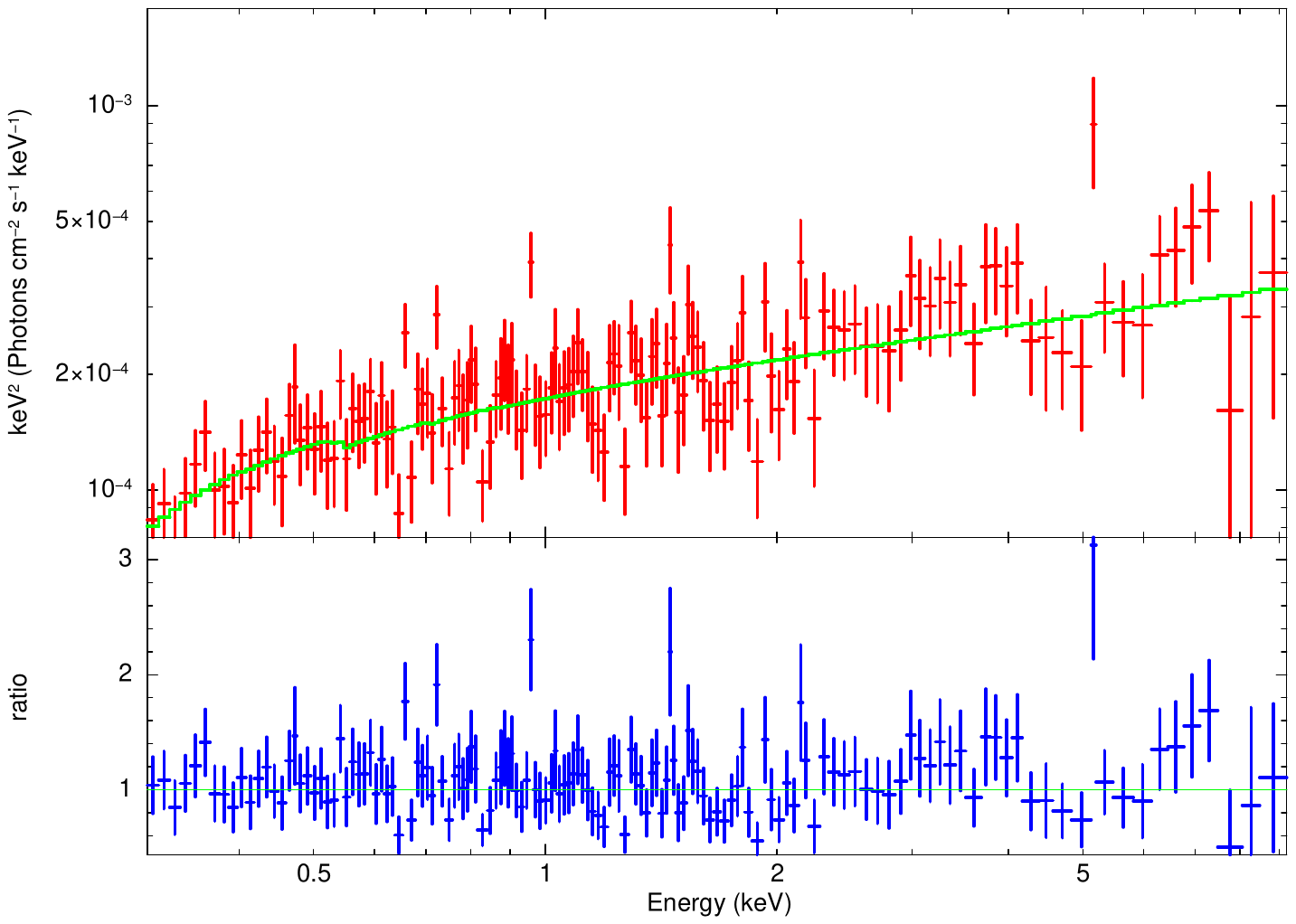}
    \includegraphics[angle=-90, width=0.47\textwidth]{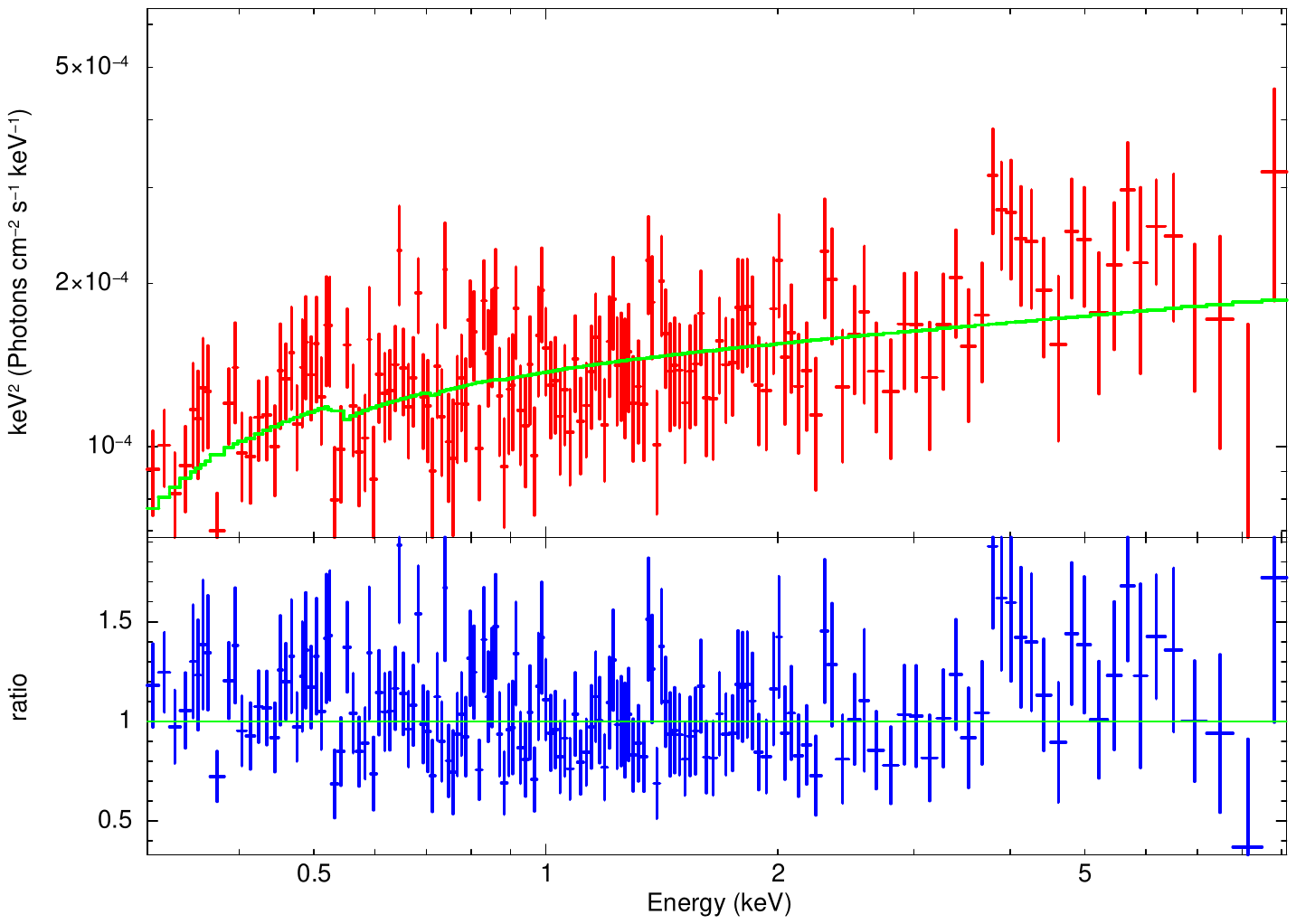}
    \includegraphics[angle=-90, width=0.47\textwidth]{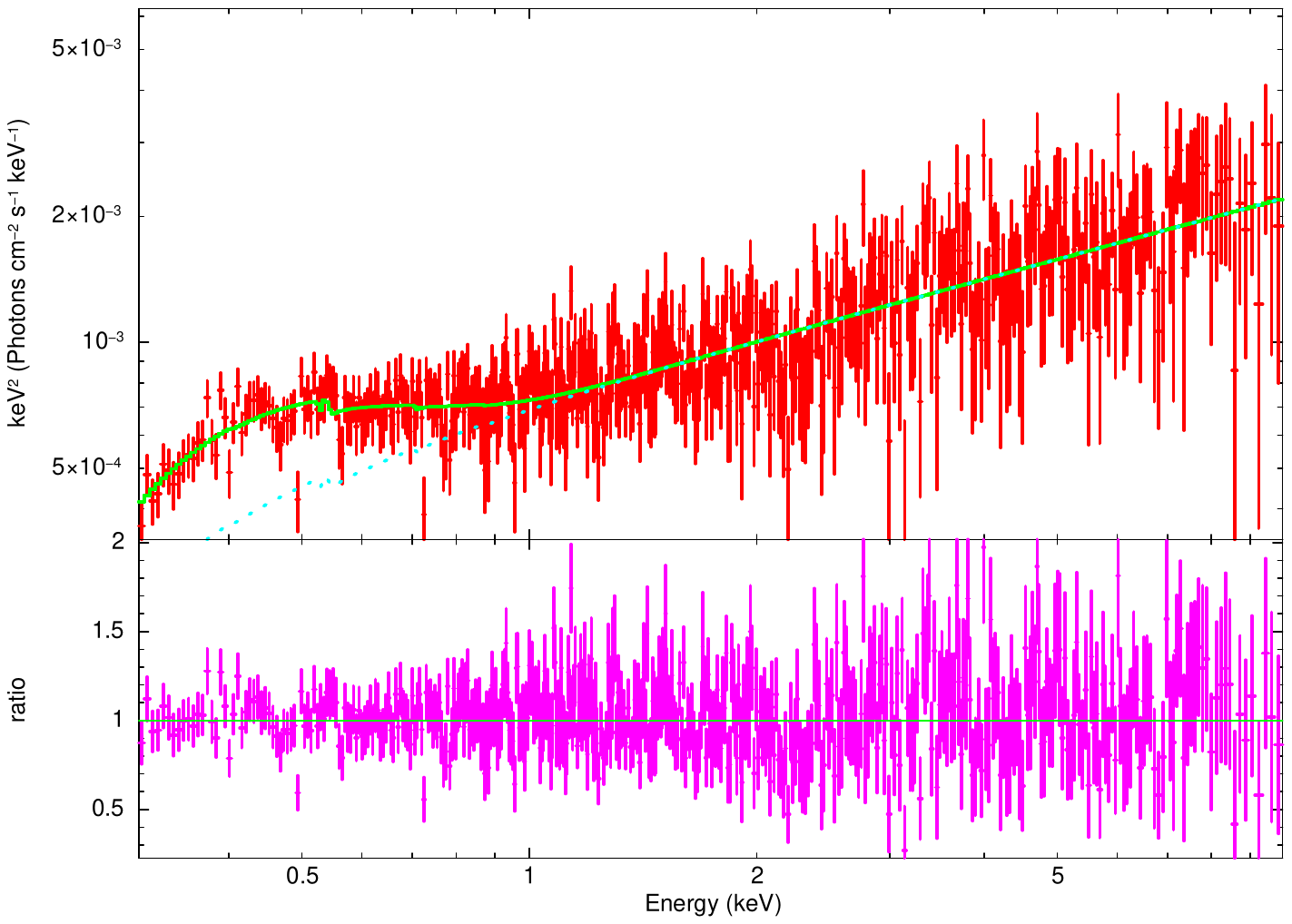}
    \caption{The XMM-Newton spectra fitted with a simple power-law model. To fit the observation during a high gamma-ray flux state disk blackbody is required (lower panel).}
    \label{fig_xmm}
\end{figure*}

During the selected period for study in this work, we found 3 observations of XMM-Newton and 2 observations of NuSTAR, which were simultaneous to two of the XMM observations. The spectra were extracted following the standard procedure for both telescopes and were fitted with a single power-law model. The best-fit parameters are shown in Table~\ref{tab:xray}, and the corresponding plots are shown in Figure~\ref{fig_xmm}. The XMM-Newton observation performed in 2019 happens to be in a low-flux state, which is also visible in Figure 1. The reduced chi-square estimated for these observations suggests that the power-law is a best-fit model. However, for observation done in June 2021, which happens to be in the high gamma-ray flux state, a single power-law does not provide a better fit; the reduced chi-square is very high. We tried different combinations to fit this XMM-Newton spectrum, such as a single log-parabola, a combination of power-law and log-parabola, a combination of black-body (bbody) and log-parabola, and a combination of power-law and black-body. Finally, we found that the power-law + bbody produces a better fit with a better-reduced chi-square. The estimated photon index is $\Gamma$ = 1.51$\pm$0.05, black-body temperature = 0.10$\pm$0.01 keV, and the chi-square is 422.57/435. This suggests that during the gamma-ray flaring state and the X-ray flaring state (Flare-1), the X-ray spectrum has some influence on black-body emission from the accretion disk, linking to a possible accretion-disk connection. A possible accretion disk connection is also suggested from the flux distribution and the PSD analysis in the next few sections. We have also produced the NuSTAR spectrum taken during Flare-1 of the gamma-ray and in the low flux state. These two spectra are used in broadband SED modeling to guide the model better in order to produce the best fit for the data and derive the physical parameters.

%The NuSTAR spectra are also fitted with a single power law and a 

\begin{table*}
\caption{ The XMM-Newton spectra are fitted with a simple power-law model. The flux is in units of 10$^{-12}$ erg cm$^{-2}$ s$^{-1}$.}
    \centering
    \renewcommand{\arraystretch}{1.2}
    \begin{tabular}{ccccccccc} \hline 
    \hline \noalign{\smallskip}
      Instruments   & Obs. Date & Obs. ID & Exposure (ks) & $\Gamma$ & $F_{unabs}$ &  $\chi^2_{r}$ & $N_H$ ($10^{22}~cm^{-2})$\\ 
      \hline \noalign{\smallskip}
      XMM-Newton &2019-05-23 & 0850390101 & 18 & 1.72$\pm$0.05 & 1.23$\pm$0.02 & 142.31/142 & 1.77 \\
      & 2019-12-26& 0850390102 & 13 & 1.88$\pm$0.05 & 0.87$\pm$0.01 & 151.55/155 & 1.77 \\
      & 2021-06-25& 0850390103 & 15 & 1.75$\pm$0.02 & 5.69$\pm$0.06 & 570.03/455 & 1.77 \\
      \hline \noalign{\smallskip}
%       NuSTAR  & 2019-05-23 & 60463037002 & 18 & 1.57$\pm$0.12 & 2.32$\pm$0.21 & 41.85/42 & 1.77  \\
%        & 2021-06-25 & 60463037004 & 17 & 1.53$\pm$0.03 & 17.59$\pm$0.46 & 150.22/171 & 1.77 \\
%\noalign{\smallskip} \hline
\label{tab:xray}
\end{tabular}
\end{table*}

%\begin{figure*}
%    \centering
%    \includegraphics[angle=-90, width=0.47\linewidth]{Tom7002.eps}
%    \includegraphics[angle=-90, width=0.47\linewidth]{Ton5997004.eps}
%    \caption{NuSTAR spectral points fitted with a single power law.}
%    \label{fig:nustar}
%\end{figure*}

\subsection{Gamma ray Emission region}
The gamma-ray light curves are produced until 6 hours of binning, which still provides a significant TS value for each data point. We used this light curve to derive the fastest variability time scale using the 
%The fastest variability time from the 6 hours binning light curve is estimated here using the 
following expression

\begin{equation}
F(t_2) = F(t_1) \cdot 2^{{(t_2 - t_1)}/{t_d}}
\end{equation}

Where \( F(t_1) \) and \( F(t_2) \) are the fluxes measured at time \(t_1\) and \(t_2\) respectively, and \(t_d\) represents doubling timescale or variability time of flux. A range of variability time is found, from a few hours to a few days. The shortest one is recorded as 2.5 hours, which is used as the fastest variability time to estimate the size of the emission region, using the following relation 
\begin{equation}
R \leq c \, t_{\text{var}} \,\delta (1 + z)^{-1},
\end{equation}
Where z=0.72 is redshift \citep{10.1111/j.1365-2966.2010.16648.x, Schneider_2010} and $\delta $ is the Doppler factor. The size of the emission region is found to be 5.0 $\times 10^{15}$ cm, for  $\delta $ = 18.2 \citep{10.1093/mnras/stad4003}. The observational constraint on the size of the emission region is important because it helps to derive the best-fit model for the broadband SED. As expected, the shortest variability time is produced by a smaller region close to the base of the jet. 

\subsection{$\gamma$-ray spectral analysis: Locating the gamma-ray emission region}
\begin{figure*}
    \centering
    \includegraphics[scale=0.5]{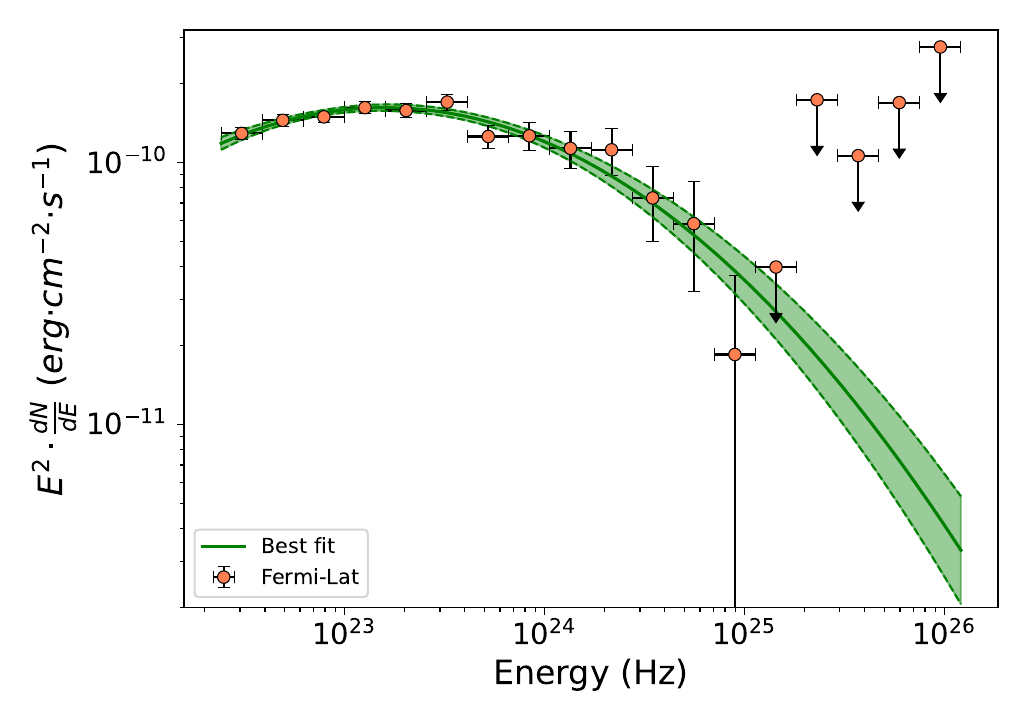}
    \includegraphics[scale=0.5]{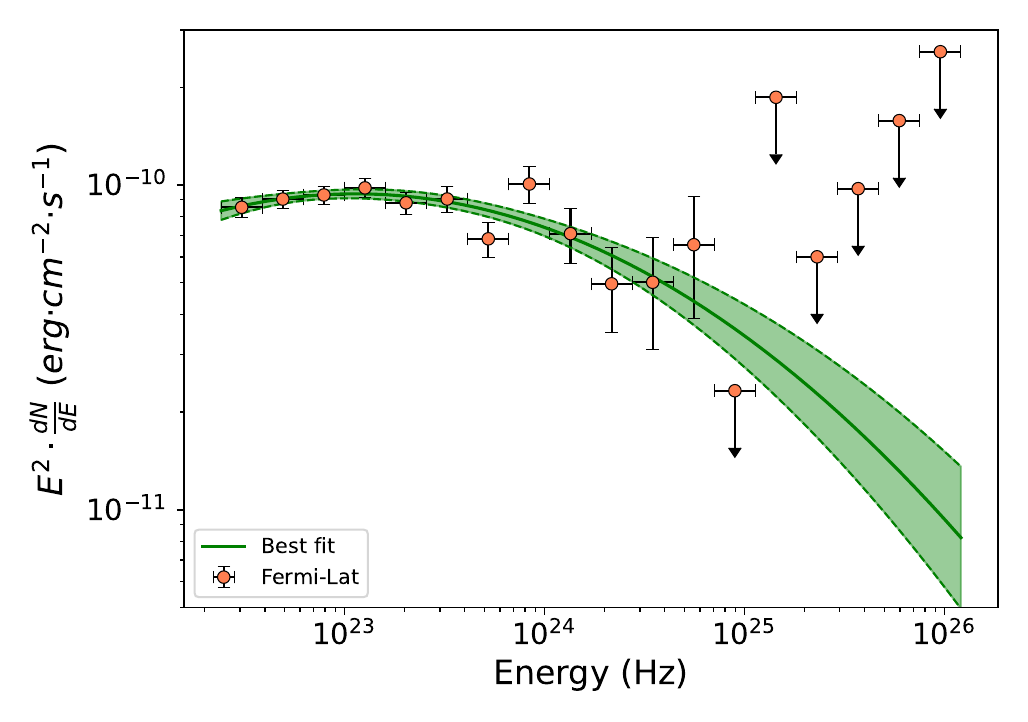}
    \includegraphics[scale=0.51]{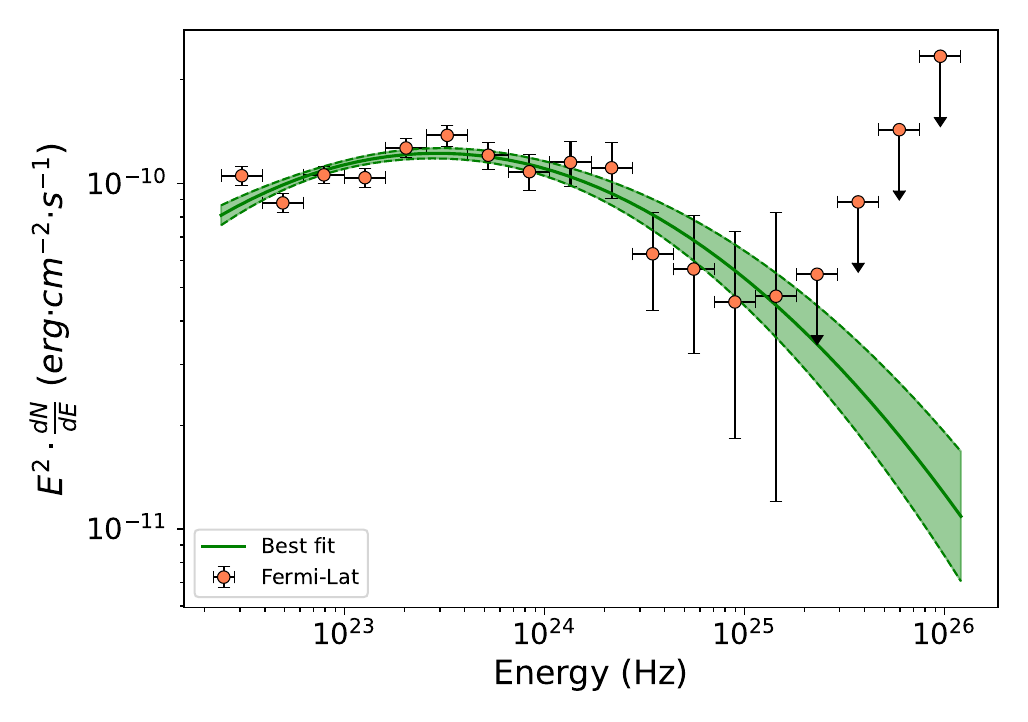}
    \includegraphics[scale=0.53]{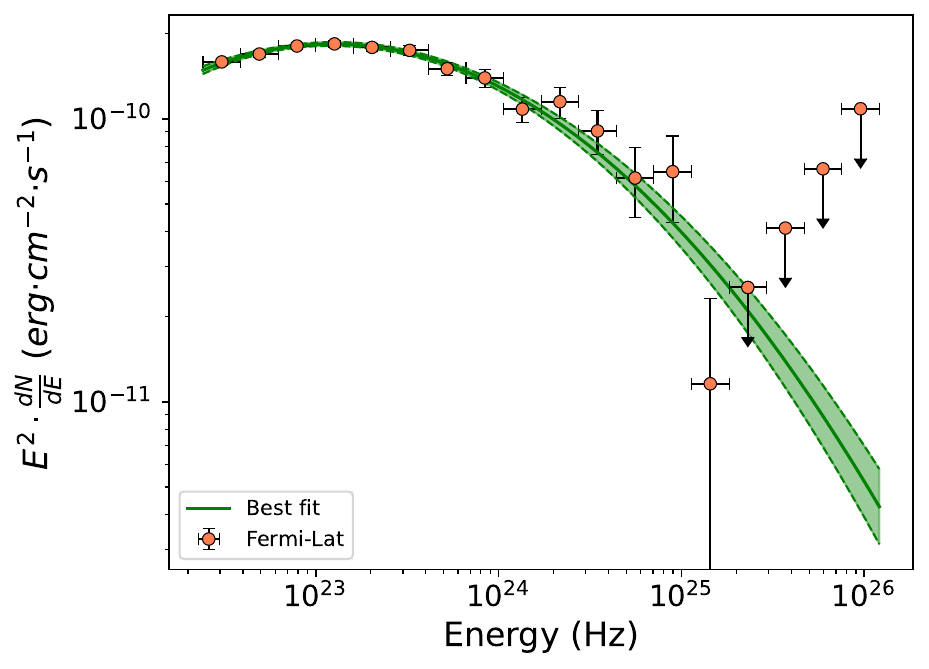}
    \caption{The gamma-ray SED derived from various states. Upper panel: Flare 1 and Quiet State. Lower panel: Flare 2 \& 3.}
    \label{fig_gamma-sed}
\end{figure*}

The location of the gamma-ray emission region is important to constrain the broadband SED, and this helps us to decide which external photon fields need to be used for the inverse-Compton scattering. To have an idea about the location of the emission region, we produced the gamma-ray spectral data points as shown in Figure~\ref{fig_gamma-sed}. As seen by the naked eye, the spectrum estimated for various flux states is curved in nature, and we fitted all the spectra with the log-parabola function to derive the best fit. We found that the log parabola fits the spectrum very well, and the best-fit parameters are shown in Table~\ref{gamma-sed}. The data points with arrows are the upper limits and have not been included in the fitting. In almost all the cases, the spectrum seems to bend around 10 GeV, suggesting that photons above 10 GeV are getting absorbed and hence reduced in number. A similar spectrum has been seen across various FSRQs and BL Lacs during the flaring events and also during the 2018 flares of Ton 599 \citep{Prince_2019}.

The presence of curvature in the gamma-ray SED plays a crucial role in constraining the location of the emission region. A curvature or break in the gamma-ray spectrum is interpreted as a signature of photon-photon absorption (pair-production) where a gamma-ray photon interacts with a low energy photon from the broad-line region (BLR), suggesting the emission region is possibly within the BLR. It has been shown that at the base of the jet, the medium is quite opaque for photons having energy above 20 GeV \citep{2006ApJ...653.1089L} and hence a curvature or break is seen in that energy range.

The curvature or the cut-off in the spectrum can happen because of other reasons as well when there is already a cut-off in the energy
distributions of the particles. The shape of the initial particle distribution injected in the jet remains unknown, and it can have a shape like power-law/broken power-law/log-parabola. 

While performing the broadband SED modeling, we keep the emission region close to BLR or within the range of the inner and outer radii of the broad-line region in order to maximise the BLR's contribution.

\begin{table*}
   \caption{Results of gamma-ray SEDs fitted with spectral type Log-Parabola(LP)}
   \label{gamma-sed}
   \centering
   \renewcommand{\arraystretch}{1.3}
  \begin{tabular}{ p{2.6cm}p{3.0cm}p{2.6cm}p{2.6cm}p{2.6cm}p{2.6cm} }
      \hline
      \hline  \noalign{\smallskip}
      \textbf{Activity} & \textbf{F} & \textbf{$\alpha$} &\textbf{$\beta$} &\textbf{$E_b$} &\textbf{TS} \\
      & [10$^{-6}$ ph cm$^{-2}$ s$^{-1}$] & & & &\\
      \noalign{\smallskip}
      \hline 
      \noalign{\smallskip}
      Flare 1& 0.851$\pm$0.164 & 1.945$\pm$0.024& 0.088$\pm$0.014 & 485.476 & 10169.736  \\
      Flare 2& 0.552$\pm$0.136& 1.879$\pm$0.029& 0.066$\pm$0.013 & 485.476 & 8379.777  \\
      Flare 3& 1.963$\pm$0.475 & 1.991$\pm$0.014 &  0.079$\pm$0.008 & 485.476 & 31278.906 \\
      Quiet Period& 0.512$\pm$0.127 & 2.003$\pm$0.030 &  0.050$\pm$0.015 & 485.476 & 5919.135 \\
      
      \hline
      \end{tabular}
\end{table*} 

\subsection{Flux distribution}
 The blazar light curve shows a range of variability on shorter to longer time scales. The cause of the variability can be accessed through the flux distribution study. The flux distribution helps us to probe the nature of variability, whether it is caused by additive or multiplicative processes. A Gaussian flux distribution leans toward an additive process, suggesting stochastic and linear variations \citep{2010LNP...794..203M}. On the other hand, if the stochastic variations are non-linear, then the log-normal flux distribution is expected, which is nothing but the Gaussian distribution of the logarithmic flux values \citep{Uttley_2005, 2010LNP...794..203M}. The log-normal distributions are very common in AGNs, Gamma-ray Burst, and the galactic X-ray binaries \citep{10.1046/j.1365-8711.2001.04496.x, 10.1093/pasj/54.5.L69, Shah_2018, 2020MNRAS.491.1934K, 2021MNRAS.502.5245P} where the emission is dominated by the accretion disk. 
 However, in the case of blazars, the situation is completely different, where most of the emission is dominated by the jets. However, in some of the studies, people have shown that the disk and jet can have some possible connection in the case of a blazar when explored systematically and carefully. In this case, the idea is that the fluctuations or variability produced in the accretion disk can somehow travel to jets and modulate the jet variability, leaving its imprint on the emissions produced in the jets. Following this scenario, it is expected that the gamma-ray emission, which is surely produced in the jet, can have a log-normal flux distribution as expected in the case of emission produced in AGN or X-ray binaries. 
 In a gamma-ray light curve of blazars \cite{2018RAA....18..141S} \& \cite{2021MNRAS.502.5245P} have found that a single log-normal flux distribution is quite prevalent.
 
 However, to counter that, there have been various suggestions that the log-normal flux distribution in the case of blazar can be produced by the mini-jets model. The mini-jets model suggests that the jets can have many mini-jets and that the total emission is the combination of these mini-jets, which leads to the multiplicative nature of the total emission and results in a log-normal flux behavior \citep{2012A&A...548A.123B}. On the other hand, \cite{Scargle_2020} argued that the log-normal flux distribution does not necessarily mean the production of a multiplicative process.

 We estimated the flux histogram of the total light curve of Ton 599 from 2020 to 2024. In the analysis, we choose only the flux data points that have TS$>$9 to account for the highly significant data points in order to achieve the correct representation.  To investigate the behavior of flux distribution, we performed the Anderson-Darling(AD) test on the logarithm of flux. We have performed the AD test for both the normal and log-normal distribution functions. The p-value for normal distribution is found to be 0.007 with AD statistics as 1.098, and for the log-normal distribution p-value is found to be 8.928$\times$ 10$^{-22}$ with AD statistics as 8.979. So, these two distributions are not suitable since the p-value is below 0.05. Next, we fit the histogram with the double Gaussian probability density functions (PDF), and the functional form is given by \cite{10.1093/mnras/stad3399} and \cite{2020MNRAS.491.1934K}.
 \begin{multline}
f(x) = \frac{a}{\sqrt{2\pi \sigma_1^2}} \exp \left( - \frac{(x - \mu_1)^2}{2 \sigma_1^2} \right)\\ 
+ \frac{(1 - a)}{\sqrt{2\pi \sigma_2^2}} \exp \left( - \frac{(x - \mu_2)^2}{2 \sigma_2^2} \right)
\end{multline}

\begin{figure}
    \centering
    \includegraphics[scale=0.37]{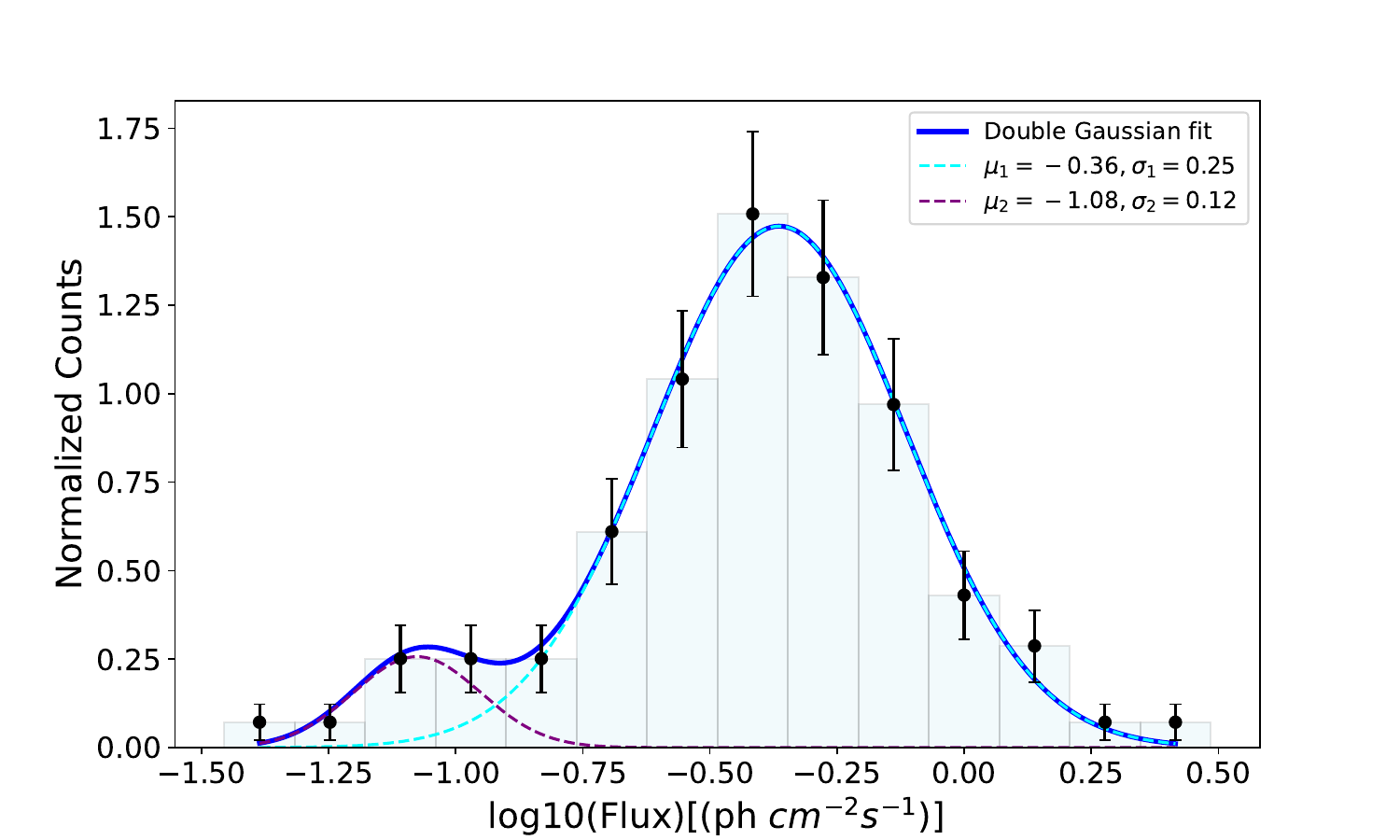}
    \caption{The figure shows double log-normal fits of the $\gamma$-ray flux histogram. The blue line represents a double log-normal fit. The parameter for fit is shown in Table~\ref{Flux_dist}.}
    \label{fig:flux-dsitribution}
\end{figure}

 The observed flux distribution is fitted with the double Gaussian function, and it is shown in Figure~\ref{fig:flux-dsitribution} and the best-fitted parameters are shown in Table~\ref{Flux_dist}. We found that a bi-model flux distribution can be well-fitted with a log-normal flux distribution, suggesting the variability of non-linear in nature. 
 As we expect, the gamma-ray emissions are produced in the jet far from an accretion disk, but if the observed jet variability is non-linear in nature or best represented with log-normal flux distribution, it suggests the possible connection between the accretion disk and the jet. It has been shown that it is quite possible, and in a few of the blazar \citep{10.1093/mnras/stad3399}, it has been shown that the variation or fluctuations produced in the accretion disk can travel to the jet and modify the jet variability accordingly.
 In literature, it also has been noticed that some blazars show a double log-normal flux distribution instead of a single log-normal distribution \citep{2018RAA....18..141S, Kushwaha_2016}. Our study concludes that there is a possibility of disk-jet coupling in this source. The nature is more clear when we look at the XMM-Newton spectra, where the thermal disk component dominates over the non-thermal emission.

\begin{table*}
   \caption{Results of the double Gaussian fit on the log-flux data for flux distribution}
   \label{Flux_dist}
   \centering
   \renewcommand{\arraystretch}{1.3}
  \begin{tabular}{ p{2.0cm}p{2.0cm}p{2.0cm}p{2.0cm}p{2.0cm}p{2.0cm}p{2.0cm} }
      \hline
      \hline  \noalign{\smallskip}
      \textbf{$\mu_1$} & \textbf{$\mu_2$} & \textbf{$\sigma_1$} &\textbf{$\sigma_2$} &\textbf{a} &\textbf{DoF} &\textbf{$\chi^2$/DoF} \\
      
      \noalign{\smallskip}
      \hline 
      \noalign{\smallskip}
      -o.36& -1.07 & 0.25& 0.12 & 0.91 & 9 & 0.71  \\

      \hline
      \end{tabular}
\end{table*} 
%\begin{figure*}
%    \centering
%    \includegraphics[scale=0.56]{hist3.png}
%    \includegraphics[scale=0.56]{hist2.png}
%    \caption{Gamma-ray flux distribution 1}
%    \label{fig:enter-label}
%\end{figure*}

\subsection{Colour-Magnitude (CM) Variation}

The colour-magnitude (CM) diagram can be measured between various optical-UV and IR filters. It can be used as a tool to study the IR-optical-UV emissions of blazars. The most common trend that has been seen among blazars is "redder-when-brighter" or "bluer-when-brighter" depending upon their types, but sometimes, a complex nature has also been seen where the trend is not clear. \cite{2012ApJ...756...13B}
have shown that mostly FSRQs follows "redder-when-brighter" whereas \cite{10.1093/pasj/63.3.327} have shown that the BL Lac in general show "bluer-when-brighter" trend.

We can also derive the optical spectral index by following the equation \citep{refId0}.
\begin{equation}
\alpha_{UB} = 0.4 \cdot \frac{U - B}{\log_{10} \left( \nu_U / \nu_B \right)},
\end{equation}
Here, (U-B) represents the colour index derived from the magnitudes in the U and B bands, while $\nu_U$ and $\nu_B$ denote the effective frequencies for these respective bands. The scaling factor in the numerator accounts for the differences between the bands \citep{1998A&A...333..231B}. 

\begin{figure*}
    \centering
    \includegraphics[scale=0.36]{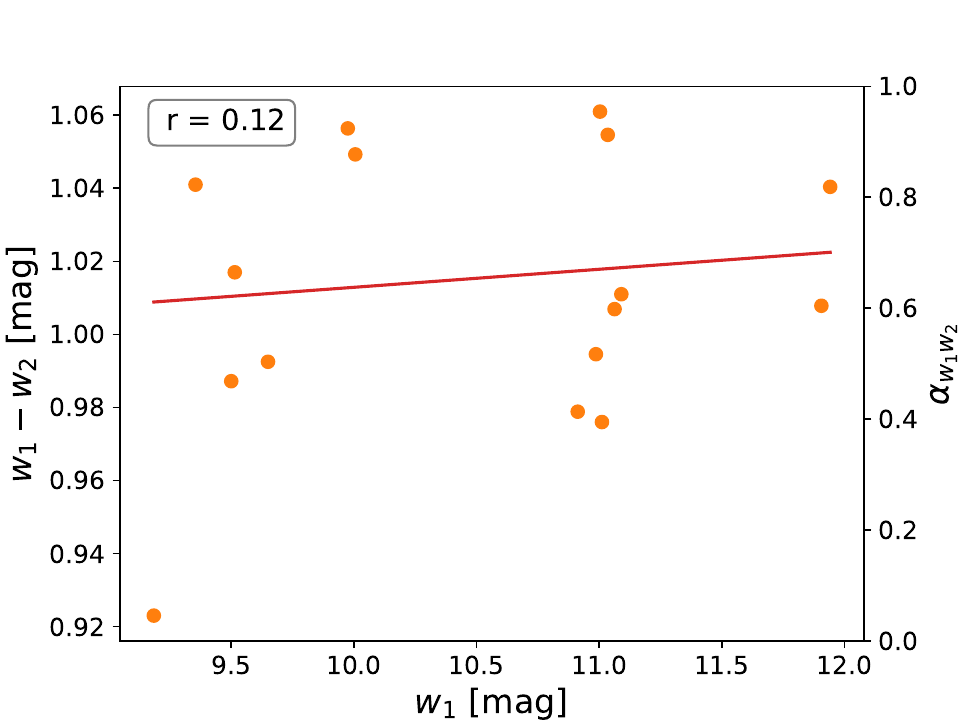}
    \includegraphics[scale=0.36]{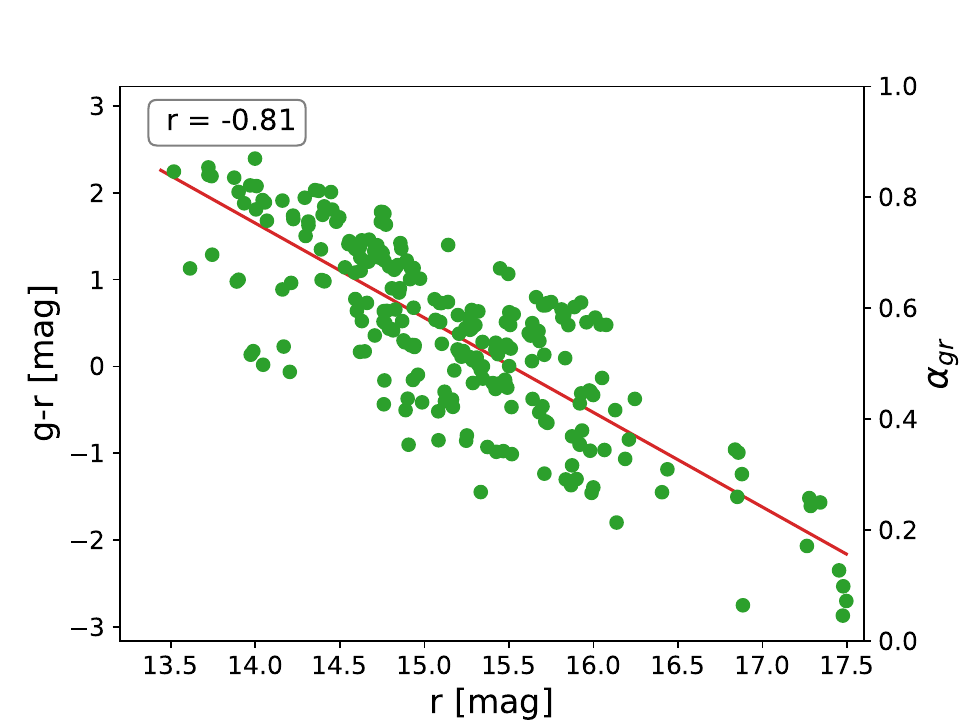}
    \includegraphics[scale=0.36]{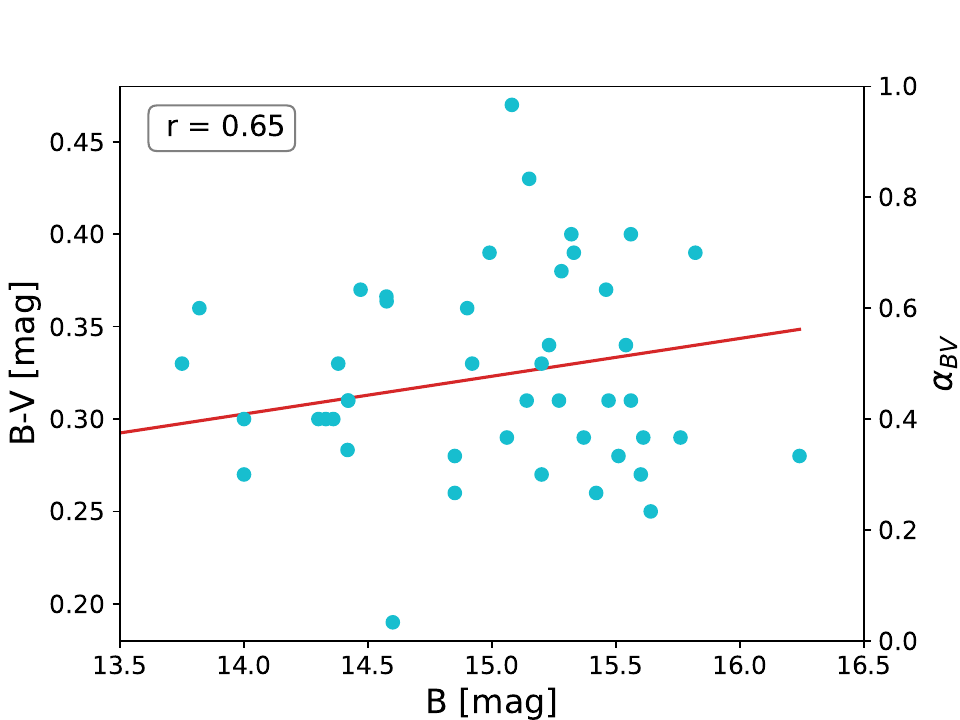}
    \includegraphics[scale=0.36]{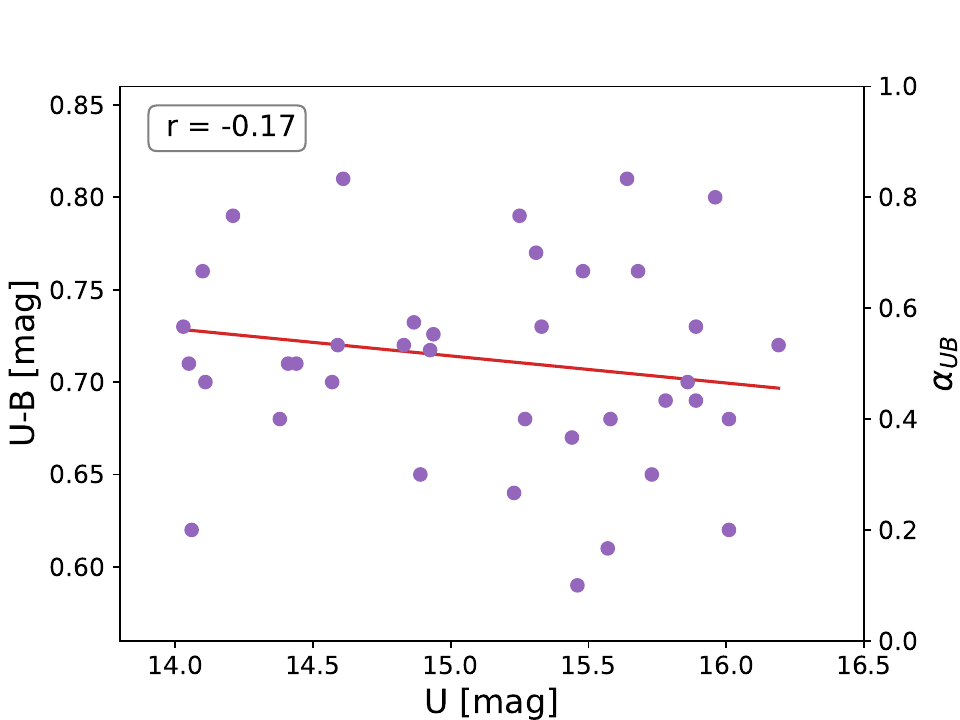}
    \includegraphics[scale=0.36]{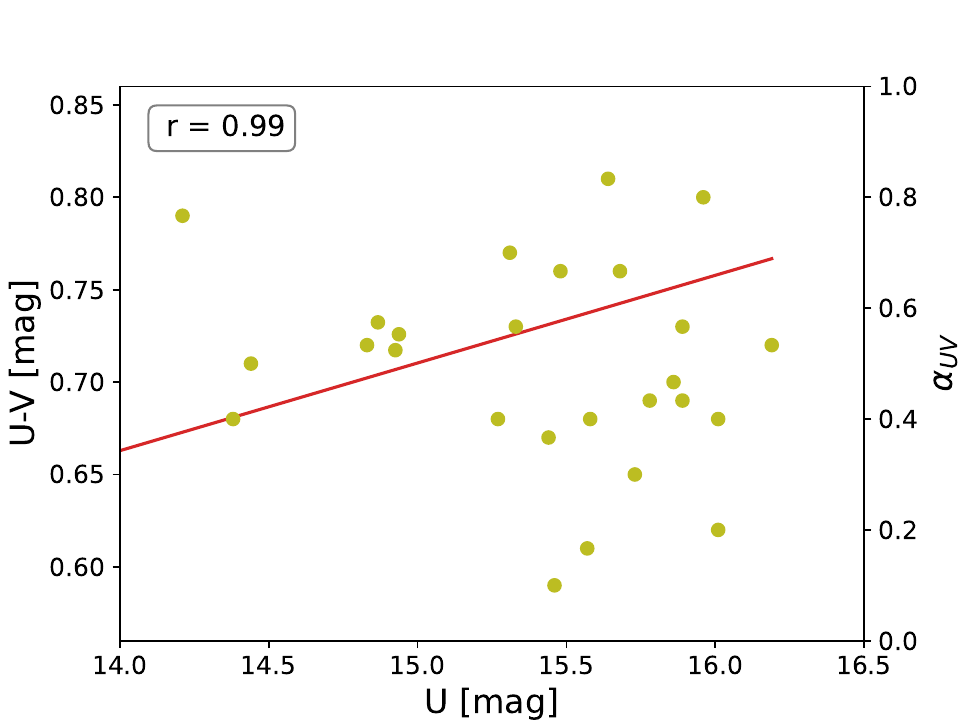}
    \caption{Colour index variations of possible IR and optical combinations from WISE, ZTF, and Swift.}
    \label{fig_CM}
\end{figure*}

In Figure~\ref{fig_CM}, we show the possible colour-index variations of combinations of WISE W1 and W2 bands, ZTF g and r bands, and Swift optical (U, B, V) bands. In the case of WISE bands, a very small number of observations are available, but they cover both the high and low flux states, as can be seen in Figure~\ref{fig_lc}, and as a result, we see a mild trend of "redder-when-brighter" with a positive correlation of $r = 0.12$. The mild trend could be because of the lack of data points in the W1 and W2 bands of WISE. Swift optical bands (B-V) and (U-V) also show similar characteristics of "redder-when-brighter" but with a much stronger positive correlation of $r = 0.65$ and $0.99$, respectively. In the case of Swift B-V, we see a mild "bluer-when-brighter" with a negative correlation coefficient of $r = -0.17$. A similar trend of "bluer-when-brighter" with a strong correlation coefficient ($r  =-0.81$) is observed in ZTF g-r. This shows that the object behaves differently in different wavebands or shows a complex nature, as also reported in \citep{10.1093/mnras/staa2622}.

In the community, the understanding is that the accretion disk emission is mostly bluer (bluer than the optical–UV from synchrotron), and the jet emission produced by the Compton scattering is mostly redder \citep{2013EPJWC..6105001G, Sarkar_2019}.

Our observation of mixed trends in optical and IR suggests that it is difficult to separate the optical-IR emission from the synchrotron and the accretion disk, and therefore, we do not see a clear trend of one behavior.

\subsection{Correlation}
The broadband correlation is an important way to understand if the multi-wavelength emissions are produced simultaneously at the same location or if they are produced at separate locations with some time delays {\citep{2014MNRAS.441.1899F}}. This information is important when we do the broadband SED modeling, which helps us to decide if one should choose a one-zone or multi-zone emission region {\citep{2018A&A...611A..44P}}. This also helps us to understand what exactly is happening inside the jets {\citep{Pushkarev_2010, Cohen_2014}}.

The broadband coverage allows us to explore the correlation among various emissions coming out of the jet. We used the traditional z-transformed discrete correlation function ($zDCF$) to find out the correlations among various wavebands and the possible time lags, if any are present there. The details about the code can be found in \citep{2013arXiv1302.1508A}. A Fortran version of the code is available online\footnote{\url{https://www.weizmann.ac.il/particle/tal/research-activities/software}}. The main reason for using zDCF over the discrete correlation function of \citep{1988ApJ...333..646E} is that it corrects several biases using equal population binning and Fisher’s z-transform. These lead to a more robust and powerful estimate of cross-correlations among sparse light curves.
If the two light curves say LC1 and LC2 are cross-correlated, then the positive time lags indicate that the LC1 light curves lead the LC2 and vice-versa. A zero time lag suggests the co-spatial origin of the emissions {\citep{Pushkarev_2010, 2013MNRAS.436.1530R, Cohen_2014, Sarkar_2019}}.

\begin{figure*}
    \centering
    \includegraphics[scale=0.25]{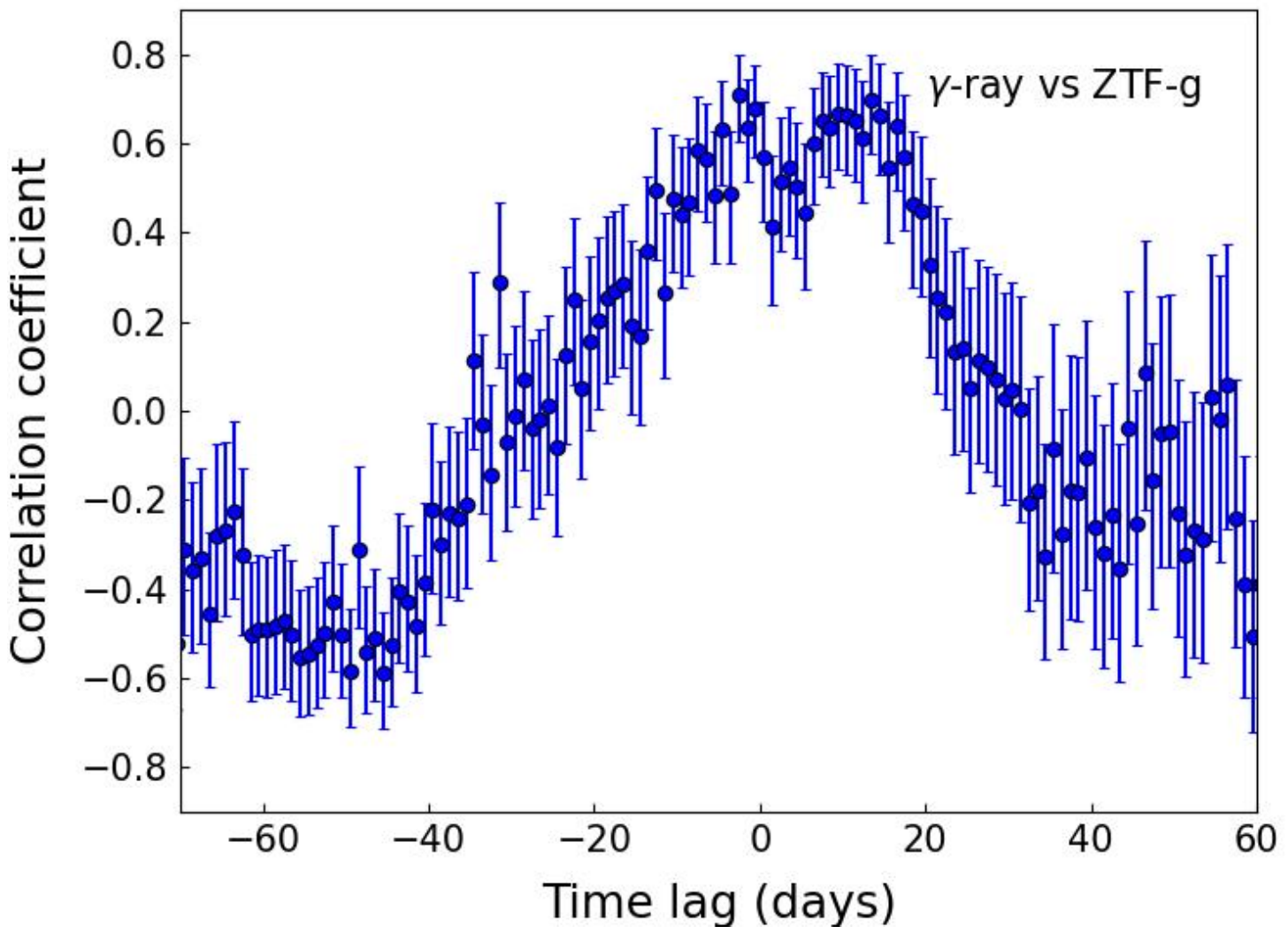}
     \includegraphics[scale=0.25]{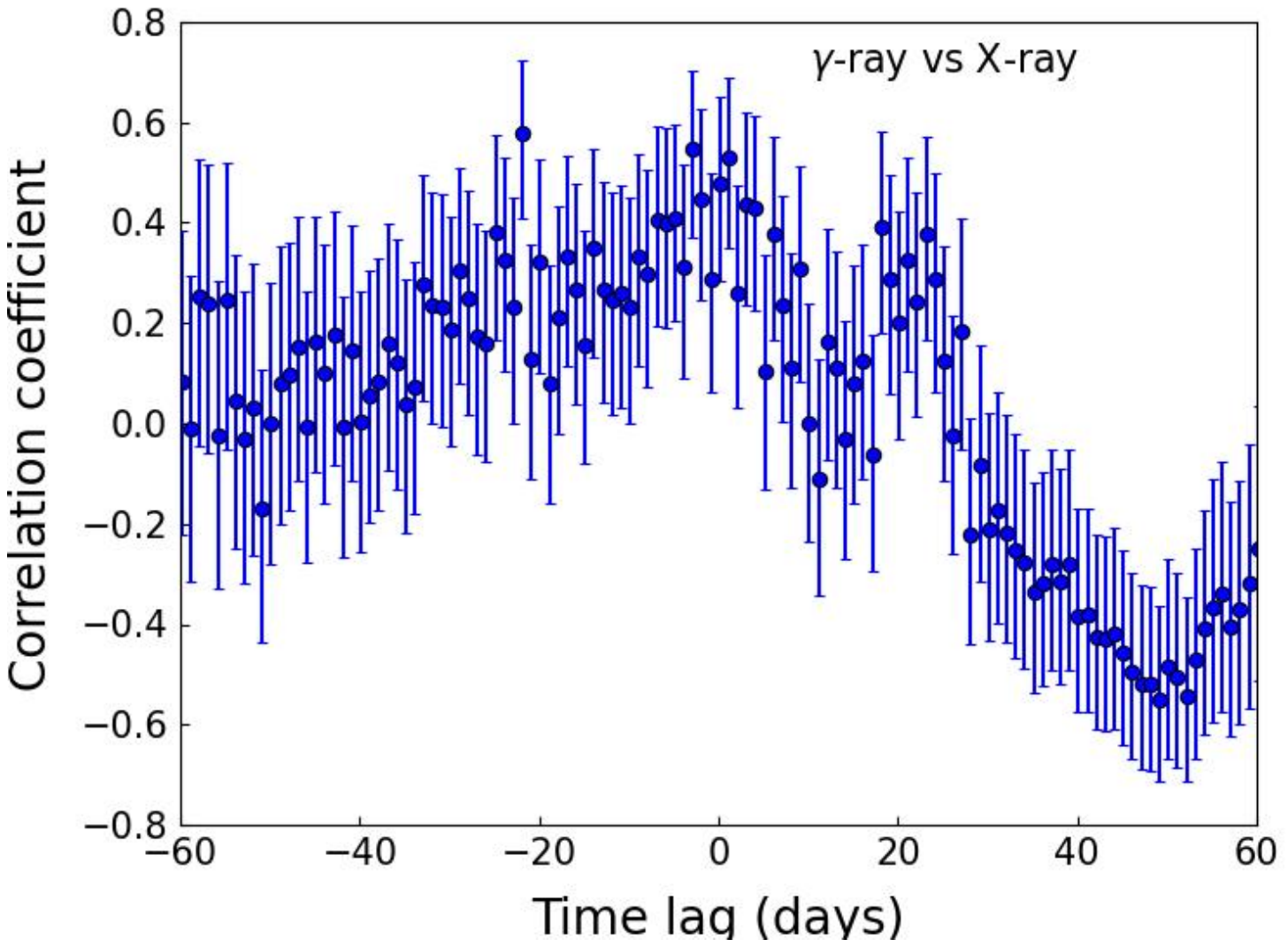}
    \caption{The cross-correlation estimated for various combinations among gamma-ray, optical, and X-ray emissions. }
    \label{fig_corr}
\end{figure*}

The gamma-ray, optical from ZTF, and the X-ray light curves are well sampled, and therefore, we choose these three frequencies for the correlation study. We focus on the total light curve rather than the individual flaring episode because we do not have a sufficient number of data to derive any significant correlations in individual flares. The correlation results are presented in Figure\ref{fig_corr}. { The correlation coefficient estimated between gamma-ray and the optical-g (ZTF) band at zero time lag is $\sim$0.7 ($\sim$70\%), suggesting the gamma-ray and optical emissions are highly correlated and are produced at the same location mostly simultaneously. The gamma-ray versus X-ray correlation shows a similar behavior with zero time lag, and the correlation coefficient is estimated as $\sim$ 0.6 ($\sim$ 60\%), suggesting gamma-ray and X-ray are also highly correlated and are produced at the same location.} Combining both results, we conclude that the gamma-ray, X-ray, and optical emissions are produced mostly simultaneously at the same location. Therefore, a single-zone emission model is sufficient to model the broadband SED. Following these arguments, we proceed with one-zone emission modeling of the broadband SED as discussed in section 3.11.
Similar results were also seen in \citep{Prince_2019} during the flare of 2018 and in the recent flare by \citet{10.1093/mnras/stad4003}.

%\begin{figure}
%    \centering  
 %   \caption{correlation}
 %   \label{fig:enter-label}
%\end{figure}

\subsection{Power spectral density}
 The variability in the source light curve can also be quantified by the power spectral density (PSD), which determines the amplitude of variation in the temporal light curve as a function of Fourier frequency or variability time scales \citep{2019ApJ...885...12R}. The PSD is important to understand the average properties of the variability, whereas the source light curve could be thought of as only a single realisation of an underlying stochastic process, as shown by \citet{vaughan2003characterizing}.

 The blazar light curve shows aperiodic flux variations across the wavebands over the short as well as the longer time scale.
The PSD can be used as a tool to derive the characteristic time scale in the aperiodic light curves, which will correspond to the variability time scale in the system, and further, it can help us to constrain the size of the emitting zone. The characteristic time scale in the system represents the breaking time in the PSD when the PSD is best fitted by the bending or broken power law rather than a simple power law.
The breaking time scale in the light curves is identified as the time scale of variability in the source or the particle cooling or escape time scales \citep{2011A&A...531A.123K, 2014ApJ...786..143S, 2014ApJ...791...21F, 2016MNRAS.458.3260C, 2017ApJ...849..138K, 2018ApJ...859L..21C, 2019ApJ...885...12R, Bhattacharyya_2020}.

In general, the observed PSD can be fitted with a single power law model, which can be defined as 
$P(\nu) \propto \nu^{-\beta}$, where $\beta$ is the slope. Depending upon the value of the slope parameter,
the earlier results suggest that the variability in high energy bands (X-ray and gamma-ray) is characterised by pink or flicker noise ($\beta=1$) \citep{Abdo_2010, Isobe_2014, Abdollahi_2017} and in lower energy (radio and optical) by damped/red-noise type processes ($\beta=2$) \citep{Max2014, Nilsson2018}.
The $\beta$ $\sim$ 0 interpreted as an uncorrelated white-noise-type
stochastic process \citep{1978ComAp...7..103P}. However, \cite{1978ComAp...7..103P} also interpreted $\beta$ $\sim$ 1 \& 2 as a flicker (or pink-noise) and damped random-walk (or
red-noise) type correlated stochastic processes, respectively.

\begin{figure}
    \centering
    \includegraphics[width=1.0\textwidth]{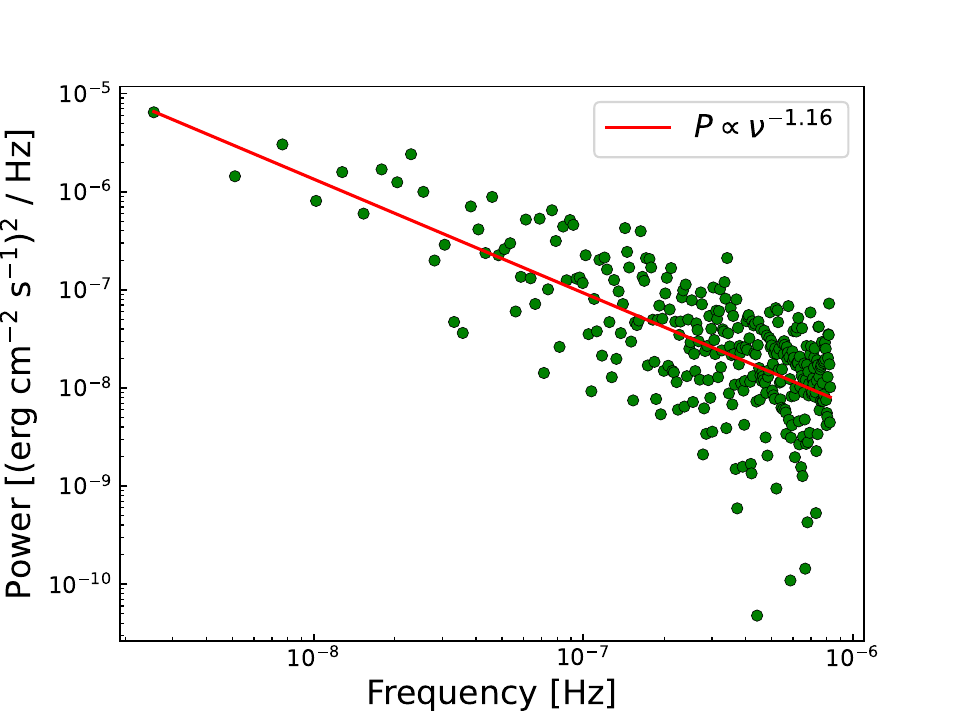}
    \caption{The Power Spectral density (PSD) derived for the total gamma-ray light curve from August 2008 to August 2024. {The continuous red line shows the best fit to the PSD with slope, $\beta=1.16$.} }
    \label{fig_psd}
\end{figure}

To probe the characteristic variability time scale and the type of variability in Ton 599, we produced the gamma-ray light curve from August 2008 to August 2024 ($\sim$ 16 years).  
We have produced the power spectrum using the discrete Fourier transform following the \citet{vaughan2003characterizing} and then fit it with a simple power law. The observed and fitted PSD is shown in Figure~\ref{fig_psd}. {The best-fitted slope ($\beta$) is estimated as $\beta=1.16$, suggesting a pink-noise kind of stochastic variability in the light curve of Ton 599. Pink noise or flickering represents that the light curve has more power in the short-term variability \citep{vaughan2003characterizing}.} We do not see any signature of curvature or break in the power spectrum,m suggesting a much longer characteristic time scale is involved in the gamma-ray variability of Ton 599 and, therefore, a much longer baseline light curve is needed to probe that. With the Fermi, which is still in operation, we hope Ton 599 will be monitored continuously for a much longer duration and possibly the break in the PSD can be observed. 

\subsection{{Disk-Jet Coupling}}
{In the literature, \citet{10.1093/mnras/179.3.433} \& \citet{10.1093/mnras/199.4.883} theorise that the jet can be collimated or launched via the extraction of rotational energy from the supermassive black hole and via the extraction of power from the accretion disk in the presence of magnetic fields, which shows that the jet and the accretion disk in some way are connected.
The observational evidence suggests that a blazar hosts mostly a thin accretion disk and strong relativistic jets. Indeed, most of the emissions that we observe in the blazar are highly dominated by the jet, and the disk emissions mostly hide behind it. However, in some cases, especially in FSRQs, people have observed strong disk emission (blackbody emission) along with high-energy gamma rays.
The concept of disc-jet coupling describes the relationship between the inflow of material through an accretion disc around a black hole and the simultaneous outflow in the form of relativistic jets \citep{Fender_2004}. This coupling plays a critical role in understanding black hole X-ray binaries, AGNs, and gamma-ray bursts. }

Observationally, the disk-jet coupling can be understood via various studies such as detecting the break in PSD \citep{2008bves.confE..14M, 2018ApJ...859L..21C} and comparing the time scale of PSD with the accretion disk time scale as done in \citet{10.1093/mnras/stad3399}. The accretion disk variability is expected to produce the log-normal flux distribution \citep{10.1046/j.1365-8711.2001.04496.x}, and hence, the gamma-ray flux distribution can be investigated to obtain the indication if the jet emission is somehow linked with the disk variability. 

We collected the XMM-Newton observation during the low and high flux states based on gamma-ray flux as marked in Figure~\ref{fig_totallc}. We analysed and fitted them using a simple power-law model. Out of the three, two spectrums are well fitted with a power law, however, the third one taken exactly during the gamma-ray flaring state required an additional model of the accretion disk to fit the data. The X-ray spectral fitting of XMM-Newton spectra is best fitted by a combination of power-law + bbody, the X-ray spectrum has the influence of black-body emission from the accretion disk, which suggests that there is a possible connection between the disk-jet which leads to the appearance of disk emission in the X-ray spectra in blazar which is very rare since the jet emission is highly dominant over the disk.

To confirm the disk-jet coupling in Ton 599, we produced the PSD using the longest gamma-ray light curve available, and we found that a single power law can fit the PSD, suggesting that an even higher baseline is required to see the break in PSD. Next, we produced the flux distribution to see if the disk variability has some influence on the gamma-ray emission. We found a bi-modal flux distribution that can be fitted with a log-normal distribution, suggesting non-linear variability in the gamma-ray emission from the jet. Given that these emissions are typically expected to originate in jets far from the accretion disk, but the observed variability is non-linear in nature and suggests that the variability produced in the accretion disk travels to the jet through an unknown process and leads to a possible connection between the accretion disk and the jet. 
It has been shown earlier that it is quite possible, and in a few of the blazar \citep{10.1093/mnras/stad3399}, it has been seen that the variation or fluctuations produced in the accretion disk can travel to the jet and modify the jet variability accordingly.
Furthermore, it has also been noticed that some blazars show a double log-normal flux distribution instead of a single log-normal distribution \citep{2018RAA....18..141S, Kushwaha_2016}. Based on the modeling of the XMM-spectrum and flux distribution study, we conclude that there is a possibility of disk-jet coupling in blazar Ton 599.

\subsection{Spectral energy distribution (SED)}
The emission mechanisms of blazars can be better understood through the modeling of their broadband spectral energy distributions (SED). We have collected broadband data from various space-based and ground-based observatories to construct the broadband SED. SED modeling provides insight into the real physical scenario happening close to the supermassive black holes. We choose to model all the identified flaring states as well as quiet or low flux states to understand which jet parameters cause the flaring events. To perform the SED modeling, we used the publicly available code {\tt JetSet}\footnote{\url{https://jetset.readthedocs.io/en/1.3.0/}}\citep{Tramacere_2009,2011ApJ...739...66T,2020ascl.soft09001T}. JETSET (version 1.3.0) fits numerical models to the data to identify the optimal parameter values that most accurately represent the observed data.  We obtained the broadband SED using the data from Fermi-LAT, Swift-XRT, Swift-UVOT, ZTF, ASAS-SN, and NuSTAR for each flare and quiet period as shown in Figure~\ref{fig_lc} except Flare 2 due to lack of observational data in that period. 
%A typical broadband SED of a blazar exhibits a characteristic two-hump structure, where the lower-energy hump is attributed to synchrotron emission and the higher-energy hump is generally attributed to photons produced through Inverse Compton scattering. The seed photons for IC scattering can originate either from within the jet, through synchrotron processes, or from external sources such as the broad-line region (BLR), dusty torus (DT), or the accretion disk. 

\begin{figure*}
    \centering
    \includegraphics[width=0.48\textwidth]{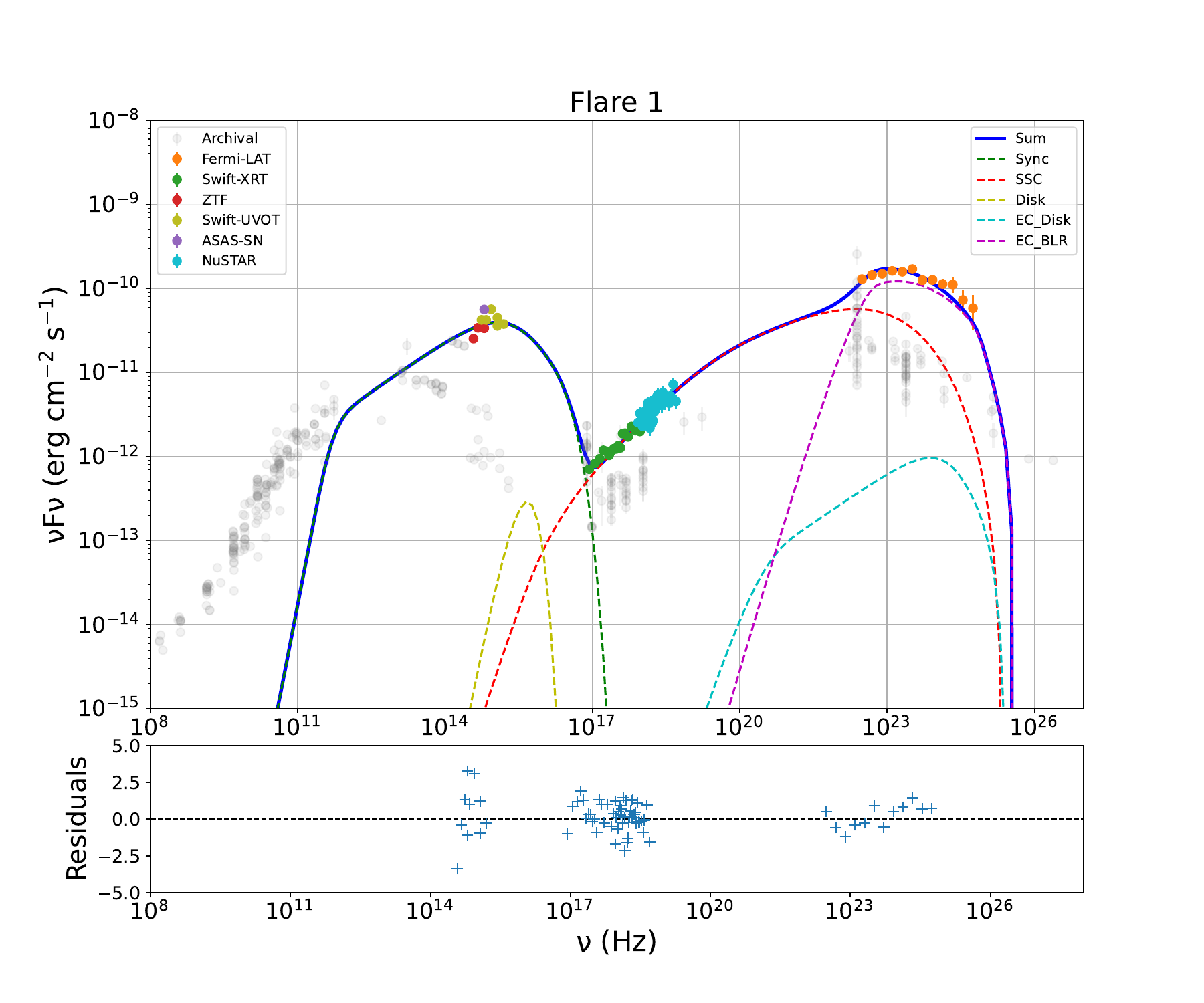}
        \includegraphics[width=0.48\textwidth]{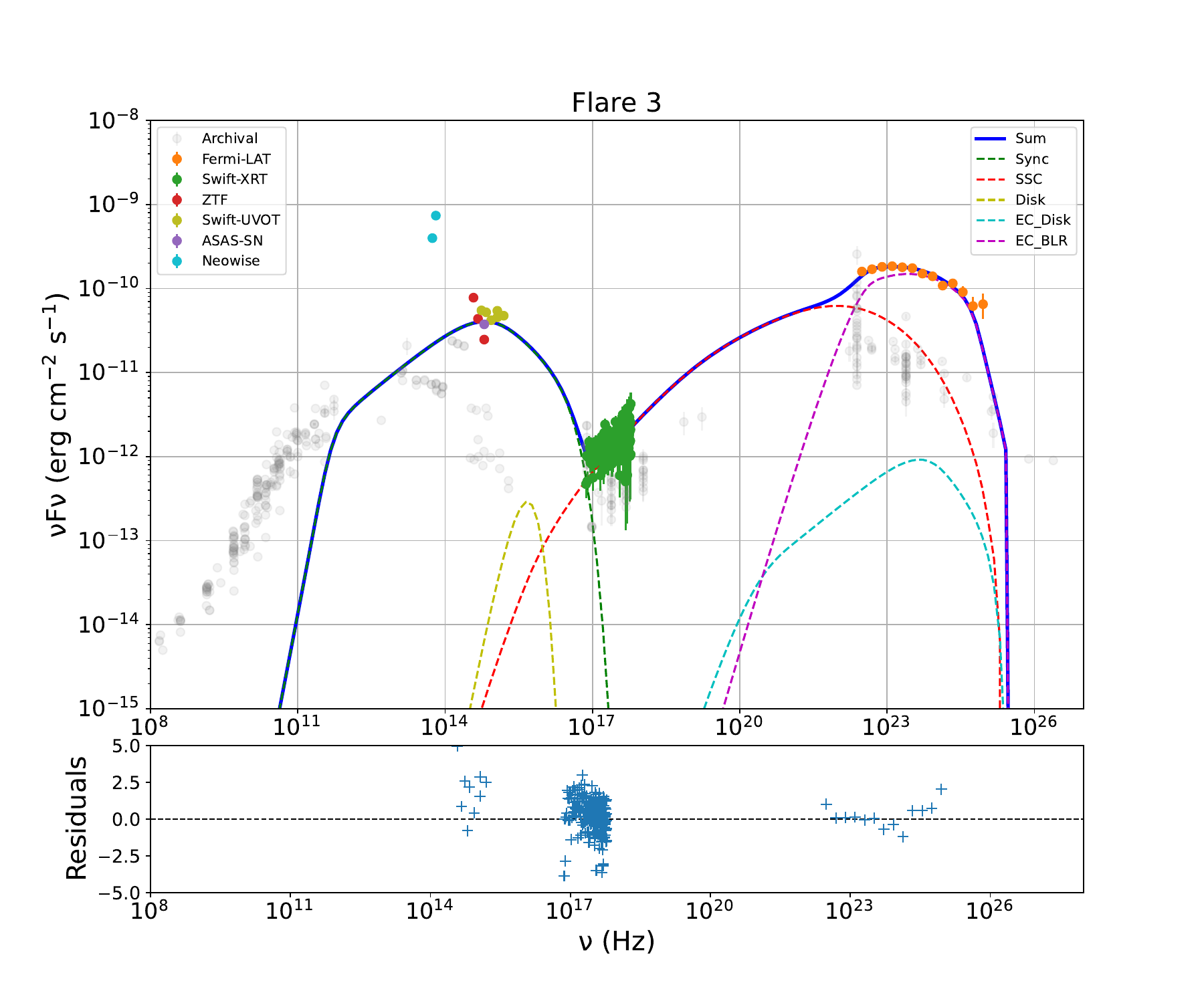}
    \includegraphics[width=0.5\textwidth]{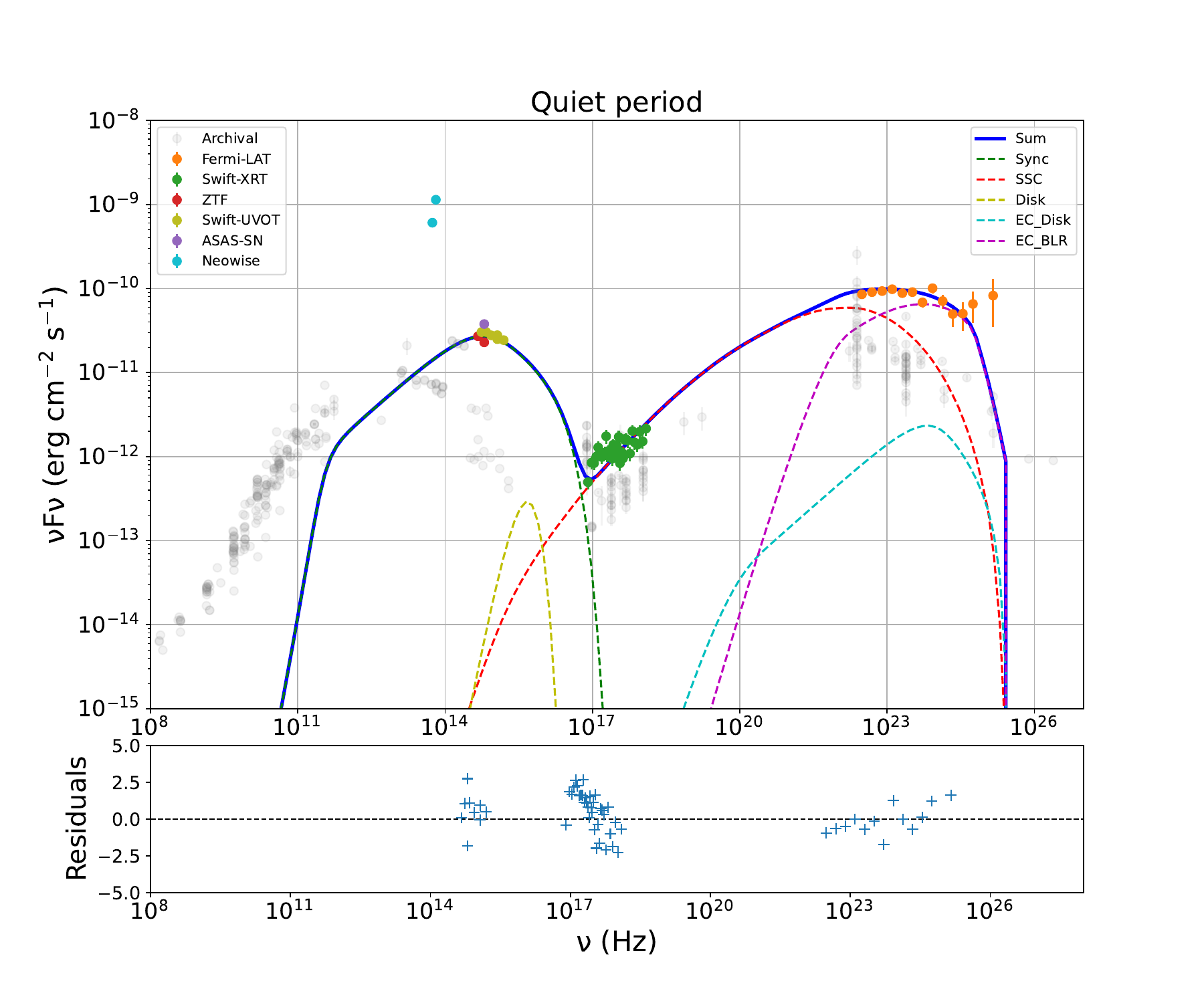}
    \caption{The broadband SEDs of Flare 1, 3 \& the quiet period fitted with one-zone leptonic model. The data and the various colourful lines are self-explanatory.}
    \label{fig_sedfig}
\end{figure*}

The broken power law (bkn) model was considered for electron distribution with a lower energy spectral slope be $p$, high energy spectral slope be $p_1$, and turn-over energy be $\gamma_{\text{break}}$. The functional form is shown in \cite{10.1093/mnras/stad4003}.
The JetSet uses this particle distribution to solve the transport equation and derive the photon flux due to synchrotron and various cases of inverse-Compton scattering.
{The JetSet has a large set of parameters in the model that will be used for modeling the SED. To reduce the number of free parameters in the model, we fixed some of the parameters such as the inner and outer radii of the BLR to standard values ($R_{BLR, in} = 2\times10^{17}$ cm; $R_{BLR, out} = 1\times10^{18}$ cm;).} { The accretion disk temperature is fixed at the standard value for the source $T_{disk}$= 1$\times10^5$ K \citep{2020MNRAS.492...72P} and the luminosity is also fixed to the standard value, $L_{disk}=1\times10^{45}$ erg/s \citep{10.1111/j.1365-2966.2009.15898.x}. The viewing angle of the jet is fixed as 2 deg \citep{2009A&A...507L..33P} and the redshift of the source to $z = 0.72$.} Through the modeling, we want to explore if the jet is particle-dominated or magnetic field-dominated. We also choose the number of cold protons in the jet that might be contributing to the production of total jet power, and for that,  we consider the ratio of cold protons to relativistic electrons to be 0.1, which is fixed during the course of fitting \citep{10.1111/j.1745-3933.2012.01280.x}. We optimize the other parameters such as magnetic field, particle distribution slopes, and energy ($\gamma_{min}$, $\gamma_{max}$), the BLR optical depth, size of the emission region, location of the emission region, and Doppler boosting to achieve a best-fit model for the broadband SED. The best-fit parameters are shown in Table~\ref{KapSou}, and the corresponding fits are shown in Figure~\ref{fig_sedfig}. 

Our modeling reveals that to fit the broadband SED, an external field is required, such as BLR and the accretion disk. These two act as external photon fields to provide a sufficient amount of photons to produce the gamma-ray emission. We noticed that higher photon densities are required during the flaring episode compared to the quiet state. We successfully fit the IR, optical, and UV spectra with the synchrotron model, and the required magnetic fields are the following: during flares 1 \& 3 the required magnetic field is almost the same (0.83 \& 0.87 Gauss respectively) and during the low flux state the magnetic field is a bit lower with the value of 0.59 Gauss. However, the magnetic field estimated in \citep{10.1093/mnras/stad4003, 10.1093/mnras/stae588} are a bit higher, with values ranging between 1.5-2.0 Gauss. Another interesting thing we noticed is that we need higher Doppler boosting to produce the bright gamma-ray flaring states (Flare 1 \& 2) compared to the low flux states. The best-fitted Doppler factor for Flare 1 \& 2 is 25 and 28 respectively, while for the quiet state, it is around 19, suggesting a sudden rise in Doppler factor can produce bright gamma-ray flaring episodes. The size of the emission region is smaller during the flaring episodes compared to the quiet state, which is expected because a smaller emission region generally produces short-term flares. 

The energy densities in the accretion disk ($U'_{Disk}$), BLR
($U'_{BLR}$), electrons ($U'_e$) and magnetic field ($U'_B$) are calculated directly by the model using JetSeT. The BLR and disk photon densities are much higher during the flaring case compared to the low flux state case, therefore supplying more photons for inverse-Compton scattering and eventually producing more gamma-ray emission. The magnetic energy density is higher in the flaring case, producing more synchrotron and SSC. The SSC plays an important role here in fitting the X-ray spectral points in the SED.

The estimated jet power in the electron and magnetic fields appears to be higher during the flaring case compared to the low flux (quiet) state, which is expected. The total jet power is estimated by summing over the electron and the magnetic power carried by the jet, and it is found to be of the order of 10$^{45}$ erg/s. Comparing this with the Eddington luminosity of the source, which is estimated as L$_{Edd}$ = 10$^{38}$(M/M$_\odot$) erg/s. In the various estimates, the mass of BH is estimated between $(0.79-3.47)\times10^8 M_{\odot}$ \citep{Xie_2004, Liu_2006}. In our case, we take
the black hole mass of Ton 599 is around 1$\times$10$^{8}$ M$_\odot$ which gives the L$_{Edd}$ = 10$^{46}$ erg/s which is much higher than the total jet power.
We conclude that the flaring events are produced by the sudden increase in the Doppler factor and the magnetic field, and the total jet power produced through this process is below the Eddington luminosity.

{\citet{2020MNRAS.492...72P} models the broadband SED of the flare of TON 599 during November 2017 using the external component (EC). Their modeling suggests the requirement of a two-component leptonic emission model to reproduce the observed broadband SED. The authors have chosen two specific periods corresponding to low and high gamma-ray flux states to compare the jet parameters. The low flux state, in their case, is well fitted with a single leptonic emission model where the GeV emission is explained by the EC-BLR. However, during the flaring state, the SED is modeled with a two-zone emission model; one is located within the BLR, and the second is outside the BLR but within the dusty torus. 
But in our study, we found that to explain the broadband SED during the flaring state, a one-zone emission model is sufficient, where the emission region is located within the broad-line region and the seed photons from the accretion disk and the BLR can act as seed photons for the external Compton process.
Their modeling suggests that during the flaring event, the jet has become magnetically dominant and, as a result, has more magnetic power than electrons. Our study suggests the jet is more particle-dominated and hence has more jet power in electrons compared to the magnetic field. The magnetic field value estimated in our case is more or less consistent with their estimate. The main difference between our results is that during the flaring state, the Doppler factor has increased significantly from $\sim$19 in the low state to $\sim$28 during the flare (i.e., especially flare 3). Other differences are in the modeling as we consider the size of the BLR as a shell rather than a single boundary.}

%\bver the electronesumming ogin{figure*}
%    \centering
%    \includegraphics[scale=0.5]{flare1_gamma_sed.pdf}
%    \includegraphics[scale=0.5]{flare2_gamma_sed.pdf}
%    \caption{Flare 2 and Flare 3}
%    \label{fig:enter-label}
%\end{figure*}

%\begin{figure}
%    \centering
%%    \includegraphics[width=0.56\textwidth]{qp_plot.pdf}
 %%   \caption{SED Quiet period}
  %  \label{fig:enter-label}
%\end{figure}
%\begin{figure*}
%    \centering
%    \includegraphics[width=0.56\textwidth]{fl3.pdf}
%    \caption{SED Flare 3}
%    \label{fig:enter-label}
%\end{figure*}

\begin{table*}
   \caption{The parameters obtained from modeling the multifrequency SEDs of flares 1, 3, and quiet period using JetSeT. The viewing angle is fixed at $\theta$=2 deg from \citep{2009A&A...507L..33P}. The parameters with (*) are fixed during modeling.}
   \label{KapSou}
   \centering
   \renewcommand{\arraystretch}{1.5}
  \begin{tabular}{ p{2cm}|p{6.5cm}|p{2cm}|p{2cm}|p{2cm} }
      \hline
      \hline  \noalign{\smallskip}
      \textbf{Symbol} & \textbf{Parameter(Units)} & \textbf{Flare1} &\textbf{Flare 3} &\textbf{Quiet Period} \\
      \noalign{\smallskip}
      \hline 
      \noalign{\smallskip}
      $\gamma_{\min}$ & Low Energy Cut-Off [$10^1$] & 72.21& 59.10 & 34.22 \\
      $\gamma_{\max}$ & High Energy Cut-Off [$10^4$] & 1.83& 1.89 & 2.12  \\
      $N$ & Emitters Density [$10^2$/cm$^3$] & 9.04& 11.57 & 8.62\\
     $\gamma_{\text{break}}$ &Turn-over Energy[$10^3$] & 5.01& 3.39 & 4.06  \\
      $p$ & Lower Energy Spectral slope & 2.19 & 2.04 & 1.46  \\
      $p_1$ & High Energy Spectral slope & 3.99 & 3.98 & 3.99 \\
      $\tau_{\text{BLR}}$ & Optical depth & 0.55& 0.70 & 0.30  \\
      $R_{\text{BLR,in}}^*$ & Inner radius of BLR [$10^{17}$ cm]  & 2 & 2 & 2 \\
      $R_{\text{BLR,out}}^*$ & Outer radius of BLR [$10^{18}$ cm]  & 1 & 1 & 1 \\
      $L_{\text{Disk}}^*$ & Disk Luminosity [$10^{45}$ erg/sec] & 1 & 1 & 1 \\
      $T_{\text{Disk}}^*$ & Disk Temperature[$10^5$ K] & 1&1&1 \\
      $R$ & Emission Region Size [$10^{15}$ cm]  & 9.28&8.54&9.71 \\
      $R_H$ & Emission Region Position [$10^{17}$cm] & 5.84&6.06&4.84 \\
      $B$ & Magnetic Field [gauss] & 0.83	&0.87&0.59 \\
      $\delta_D$ &Jet Bulkfactor & 25.31&28.33&19.47 \\
      $ \theta^* $ & Jet viewing angle & 2	&2&2 \\
      $z_{\text{cosm}}^*$ & Redshift & 0.72 & 0.72 & 0.72 \\
      $N_{\text{H\_cold\_to\_rel\_e}}^*$ & Cold proton to relativistic electron ratio & 0.1 & 0.1 & 0.1 \\

      \noalign{\smallskip}
      \hline
      \hline
      \multicolumn{5}{c}{\textbf{Energy Densities}} \\
      
      \hline 
      \noalign{\smallskip}
      $U_{\text{BLR}}$ & BLR Energy Density [$10^{-2} \, \text{erg/cm}^3$] & 252.72 & 376.61 & 110.64  \\ 
      $U_{\text{Disk}}$ & Disk Energy Density [$10^{-2} \, \text{erg/cm}^3$] & 8.51 & 8.21 & 0.15 \\
      $U_{\text{e}}$ & Electron Energy Density[$10^{-1} \, \text{erg/cm}^3$] & 1.98 & 2.41 & 2.61 \\
      $U_{\text{B}}$ & Magnetic Field Energy Density[$10^{-2} \, \text{erg/cm}^3$] & 2.75 & 3.07 & 1.43 \\
     \noalign{\smallskip}
     \hline 
     \hline
     \multicolumn{5}{c}{\textbf{Jet Luminosity}} \\
     \hline \noalign{\smallskip}
     $L_{\text{e}}$ & Jet Lepton Luminosity[$10^{45} \, \text{erg/cm}$] & 1.03 & 1.33 & 0.88 \\
     $L_{\text{B}}$ & Jet Magnetic Field Luminosity[$10^{44} \, \text{erg/cm}$] & 1.43 & 1.69 & 0.48 \\
     $L_{\text{Jet}}$ & Total Jet Luminosity[$10^{45} \, \text{erg/cm}$] & 1.17 & 1.78 & 0.92 \\
     \noalign{\smallskip}
     \hline 
     \end{tabular}
\end{table*}  

\section{Summary}
 The Bayesian block methodology turned out to be the best way to characterise the light curve into various high and low flux states.
 We identified three bright gamma-ray flares during the years 2019 to 2024.
The rise and decay time of flares are found to be in the range of 1 - 7 hours.
The F$_{\text{var}}$ shows a double hump structure resembling the broadband SED.
The XMM-Newton spectra taken during the flaring episode reveal the presence of a thermal black-body component, suggesting the possible contribution of the accretion disk in the total jet emission.
The gamma-ray spectral points show a clear break around 10 GeV, suggesting the emission region to be located within the BLR.
 A double-log normal flux distribution is found, which is rarely seen in blazars.
The gamma-ray, optical, and X-ray emissions were found to be highly correlated with zero time lag, suggesting their co-spatial origin.
 A total of 16 years of Fermi data are used to derive the PSD, which is best fitted with a single power law, suggesting that a long baseline is required to probe the characteristic time scale.
 One-zone leptonic model is successfully used to fit the broadband data, and the finding is that the high flux state is most probably caused by the increase in the magnetic field and the rise in the Doppler factor.
 The results of XMM-Newton spectral modeling, along with the flux distribution, suggest there is a possible disk-jet coupling in this source.

\section{Acknowledgments}
We thank the referee for their constructive suggestions, which have helped to improve the manuscript.
This research makes use of the publicly available data from Fermi-LAT obtained from the FSSC data server and distributed by NASA Goddard Space Flight Center (GSFC). This work made use of data supplied by the UK Swift Science Data Centre at the University of Leicester. The data, software, and web tools obtained from NASA’s High Energy Astrophysics Science Archive Research Center (HEASARC), a service of GSFC, are used in this work. This work also makes use of publicly available data from NuSTAR and XMM-Newton obtained from the HEASARC Archive.

%\endnote in some journals will behave like \footnote; and \printendnotes will not output anything. 
\printendnotes

%\printbibliography
\bibliography{ref}

\end{document}